\documentclass[prd,twocolumn,twoside,preprintnumbers,superscriptaddress,nofootinbib]{revtex4}

\usepackage{amsmath,slashed}
\usepackage{graphicx,graphics,color}
\usepackage{dcolumn}
\usepackage[hyperfootnotes=false]{hyperref}
\usepackage{xspace}
\usepackage{enumerate}
\usepackage{varwidth}

\usepackage{physics}
\usepackage{amssymb}
\usepackage{mathtools}
\usepackage{dsfont}
\usepackage{multirow}

\hypersetup{
    colorlinks=true,
    linkcolor=blue,
    citecolor=blue
}

\newcommand{\beq}{\begin{equation}}
\newcommand{\eeq}{\end{equation}}

\newcommand{\GeV}{\,\text{GeV}}

\newcommand{\Mr}{M_\rho}

\makeatletter
\newcommand{\Cr}[2]{\@ifmtarg{#2}{\mathcal{C}_{#1}}{\mathcal{C}_{#1}\big[#2\big]}}
\makeatother

\allowdisplaybreaks[1]

\begin{document}

\preprint{PSI-PR-24-27, ZU-TH 61/24}

\title{Complete dispersive evaluation of the hadronic light-by-light
    contribution\\[1mm] to muon $\boldsymbol{g-2}$}

\author{Martin Hoferichter}
\affiliation{Albert Einstein Center for Fundamental Physics, Institute for Theoretical Physics, University of Bern, Sidlerstrasse 5, 3012 Bern, Switzerland}
\author{Peter Stoffer}
\affiliation{Physik-Institut, Universit\"at Z\"urich, Winterthurerstrasse 190, 8057 Z\"urich, Switzerland}
\affiliation{PSI Center for Neutron and Muon Sciences, 5232 Villigen PSI, Switzerland}
\author{Maximilian Zillinger}
\affiliation{Albert Einstein Center for Fundamental Physics, Institute for Theoretical Physics, University of Bern, Sidlerstrasse 5, 3012 Bern, Switzerland}

\begin{abstract} 
Hadronic light-by-light (HLbL) scattering  defines one of the critical contributions in the Standard-Model prediction of the anomalous magnetic moment of the muon. In this Letter, we present a complete evaluation using a dispersive formalism, in which the HLbL tensor is reconstructed from its discontinuities, expressed in terms of simpler hadronic matrix elements that can be extracted from experiment. Profiting from recent developments in the determination of axial-vector transition form factors, short-distance constraints for the HLbL tensor, and the vector--vector--axial-vector correlator, we obtain
 $a_\mu^\text{HLbL}=101.9(7.9)\times 10^{-11}$, which meets the precision requirements set by the final result of the Fermilab experiment.
\end{abstract}

\maketitle

\emph{Introduction}---Lepton anomalous magnetic moments, $a_\ell =(g-2)_\ell/2$, have long served as precision tests of the Standard Model, going back to Schwinger's seminal calculation~\cite{Schwinger:1948iu} and its experimental verification~\cite{Kusch:1948mvb}. The case of the muon, $\ell=\mu$, is particularly interesting given that potential contributions beyond the Standard Model are enhanced by the lepton mass, and the experimental world average now stands at
\beq
a_\mu^\text{exp}=116\,592\,059(22)\times 10^{-11},
\eeq
completely dominated by the Fermilab experiment~\cite{Muong-2:2023cdq,Muong-2:2024hpx}. The precision will be improved further with the upcoming final result including Runs 4+5+6, with a projected precision around $\Delta a_\mu^\text{exp}\simeq 13\times 10^{-11}$ that would even surpass the original design sensitivity~\cite{Muong-2:2015xgu}.

\begin{figure}[t]
     \centering
     \includegraphics[width=0.25\linewidth]{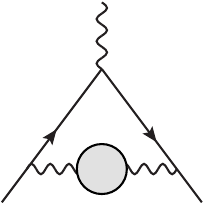}\qquad
     \includegraphics[width=0.25\linewidth]{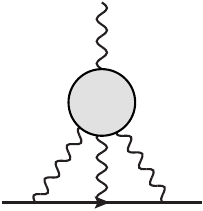}
     \caption{HVP (left) and HLbL (right) contributions to $a_\mu$. The solid lines denote  muons, the wiggly lines photons, and the gray blobs refer to the hadronic two- and four-point functions, respectively.}
     \label{fig:HLbL}
 \end{figure}

Such a precision presents a formidable challenge to theory~\cite{Aoyama:2020ynm}.
While QED~\cite{Aoyama:2020ynm,Aoyama:2012wk,Aoyama:2019ryr} and electroweak~\cite{Czarnecki:2002nt,Gnendiger:2013pva} contributions are well under control, the same is not true for
the hadronic corrections, hadronic vacuum polarization (HVP)~\cite{Davier:2017zfy,Keshavarzi:2018mgv,Colangelo:2018mtw,Hoferichter:2019gzf,Davier:2019can,Keshavarzi:2019abf,Hoid:2020xjs}
and hadronic light-by-light (HLbL) scattering~\cite{Melnikov:2003xd,Masjuan:2017tvw,Colangelo:2017qdm,Colangelo:2017fiz,Hoferichter:2018dmo,Hoferichter:2018kwz,Gerardin:2019vio,Bijnens:2019ghy,Colangelo:2019lpu,Colangelo:2019uex,Pauk:2014rta,Danilkin:2016hnh,Jegerlehner:2017gek,Knecht:2018sci,Eichmann:2019bqf,Roig:2019reh}, see Fig.~\ref{fig:HLbL} (uncertainties due to higher-order iterations are again sufficiently small~\cite{Calmet:1976kd,Kurz:2014wya,Colangelo:2014qya,Hoferichter:2021wyj}). This is particularly true for the HVP contribution, in which case tensions
both within data-driven approaches~\cite{Davier:2017zfy,Keshavarzi:2018mgv,Colangelo:2018mtw,Hoferichter:2019gzf,Davier:2019can,Keshavarzi:2019abf,Hoid:2020xjs,Crivellin:2020zul,Keshavarzi:2020bfy,Malaescu:2020zuc,Colangelo:2020lcg,Stamen:2022uqh,Colangelo:2022vok,Colangelo:2022prz,Hoferichter:2023sli,Hoferichter:2023bjm,Stoffer:2023gba,Davier:2023fpl,CMD-3:2023alj,CMD-3:2023rfe,Leplumey:2025kvv} and with lattice QCD~\cite{Borsanyi:2020mff,Ce:2022kxy,ExtendedTwistedMass:2022jpw,FermilabLatticeHPQCD:2023jof,RBC:2023pvn,Boccaletti:2024guq,Blum:2024drk,Djukanovic:2024cmq,Bazavov:2024eou} currently limit the precision, but community-wide efforts are under way to resolve the situation~\cite{Colangelo:2022jxc}, e.g., by revisiting the role of   radiative corrections~\cite{Campanario:2019mjh,Ignatov:2022iou,Colangelo:2022lzg,Monnard:2021pvm,Abbiendi:2022liz,BaBar:2023xiy,Aliberti:2024fpq}. In the meantime, it is important to emphasize that also the HLbL contribution, $a_\mu^\text{HLbL} = 92(19) \times 10^{-11}$~\cite{Aoyama:2020ynm,Melnikov:2003xd,Masjuan:2017tvw,Colangelo:2017qdm,Colangelo:2017fiz,Hoferichter:2018dmo,Hoferichter:2018kwz,Gerardin:2019vio,Bijnens:2019ghy,Colangelo:2019lpu,Colangelo:2019uex,Pauk:2014rta,Danilkin:2016hnh,Jegerlehner:2017gek,Knecht:2018sci,Eichmann:2019bqf,Roig:2019reh}, is not known with the precision mandated by the final result of the Fermilab experiment, and improvement by at least a factor two in precision is needed. In this Letter, we present a data-driven HLbL evaluation that meets these requirements.

To this end, we employ a dispersive approach~\cite{Colangelo:2014dfa,Colangelo:2014pva,Colangelo:2015ama,Colangelo:2017fiz,Hoferichter:2013ama}, whose basic idea amounts to reconstructing the entire HLbL tensor from its discontinuities via a dispersion relation, using the fact that these discontinuities are determined by simpler matrix elements that can be extracted from experiment. Such a dispersive approach in terms of exclusive hadronic intermediate states applies only up to a certain energy, above which a matching to short-distance constraints (SDCs) needs to be performed. In this Letter, we provide a complete dispersive evaluation, profiting from three recent developments: (i) the transition form factors (TFFs) for the axial-vector resonances, the largest intermediate states not evaluated dispersively so far, were analyzed in a global fit to the available data on the $f_1$~\cite{Zanke:2021wiq,Hoferichter:2023tgp}, their asymptotic behavior was studied using the light-cone expansion~\cite{Hoferichter:2020lap}, and an optimized basis for the HLbL tensor was derived to make the absence of kinematic singularities manifest~\cite{Hoferichter:2024fsj}; (ii) subleading corrections to the SDCs were calculated~\cite{Bijnens:2020xnl,Bijnens:2021jqo,Bijnens:2022itw,Bijnens:2024jgh}; (iii) the vector--vector--axial-vector ($VVA$) correlator, needed as input for the operator product expansion (OPE) in part of the parameter space, was studied in a dispersive approach, providing also a dispersive reconstruction of the singly-virtual $a_1$ TFF~\cite{Ludtke:2024ase}. In the following, we lay out the essential steps in this program, referring to Ref.~\cite{Hoferichter:2024bae} for a detailed description.

\begin{figure}[t]
     \centering
     \includegraphics[width=\linewidth]{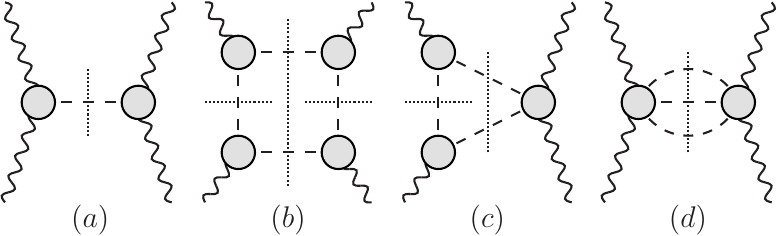}
     \caption{Unitarity diagrams in a dispersive approach to HLbL scattering: $(a)$ pseudoscalar poles, $(b)$ pion/kaon box, $(c)$ example of rescattering corrections, $(d)$ multi-meson intermediate states. Dashed lines refer to mesons, wiggly lines to photons, gray blobs to hadronic matrix elements, and dotted lines indicate that the respective states are taken on-shell.}
     \label{fig:disp}
 \end{figure}

\emph{Dispersive formalism}---The starting point of the dispersive approach is a Bardeen--Tung--Tarrach~\cite{Bardeen:1968ebo,Tarrach:1975tu} decomposition of the HLbL tensor~\cite{Colangelo:2015ama} into scalar functions $\Pi_i$
\beq
\label{decomposition}
\Pi^{\mu\nu\lambda\sigma} = \sum_{i=1}^{54} T_i^{\mu\nu\lambda\sigma} \Pi_i,
\eeq
where the Lorentz structures $T_i^{\mu\nu\lambda\sigma}$ for the photon--photon scattering process
\beq
\label{photon-photon}
\gamma^*(q_1,\mu)\gamma^*(q_2,\nu)\to\gamma^*(-q_3,\lambda)\gamma^*(q_4,\sigma)
\eeq
are given in Ref.~\cite{Colangelo:2017fiz}. The master formula for the HLbL contribution can be written in the form
\begin{align}	\label{eq:MasterFormulaPolarCoord}
	a_\mu^\text{HLbL} &= \frac{\alpha^3}{432\pi^2} \int_0^\infty d\Sigma\, \Sigma^3 \int_0^1 dr\, r\sqrt{1-r^2} \int_0^{2\pi} d\phi \notag\\
	&\qquad \times\sum_{i=1}^{12} T_i(\Sigma,r,\phi) \bar\Pi_i(q_1^2,q_2^2,q_3^2),
\end{align}
with known kernel functions $T_i(\Sigma,r,\phi)$~\cite{Colangelo:2017fiz} and photon virtualities $Q_i^2=-q_i^2$ parameterized as~\cite{Eichmann:2015nra}
\begin{align}
\label{Qi}
		Q_{1/2}^2 &= \frac{\Sigma}{3} \left( 1 - \frac{r}{2} \cos\phi \mp \frac{r}{2}\sqrt{3} \sin\phi \right), \notag\\
		Q_3^2 &= \frac{\Sigma}{3} \left( 1 + r \cos\phi \right).
\end{align}
The $12$ scalar functions $\bar\Pi_i(q_1^2,q_2^2,q_3^2)$ derive from the original $54$ $\Pi_i$ and represent the dynamical content of the theory. In this Letter, we follow the original approach from Refs.~\cite{Colangelo:2015ama,Colangelo:2017fiz} and consider dispersion relations in four-point kinematics, i.e., for the Mandelstam variables of the scattering process~\eqref{photon-photon} with fixed photon virtualities, in contrast to an alternative approach in triangle kinematics~\cite{Ludtke:2023hvz}, in which case the external photon is taken soft prior to setting up the dispersion relations. Both approaches are equivalent if the infinite tower of intermediate states is resummed, the main advantage of four-point kinematics being that the dependence of pole contributions on the photon virtualities  is automatically resummed into TFFs, the disadvantage being that even in the optimized basis from Ref.~\cite{Hoferichter:2024fsj} the cancellation of kinematic singularities for spin $\geq 2$ is no longer manifest. Due to the different truncation of intermediate states a comparison of the two approaches should in the future enable the best control over remaining uncertainties in the matching between hadronic and asymptotic continuum contributions. For the numerical analysis, we will show results separately for $\bar\Pi_{1,2}$ and $\bar\Pi_{3\text{--}12}$, motivated by the fact that the dominant pseudoscalar poles due to $\pi^0$, $\eta$, $\eta'$ only contribute to the former and that, in the OPE limit discussed below, the two classes map onto longitudinal and transverse contributions, respectively.

\emph{Exclusive hadronic states}---The dominant contributions arise from the light pseudoscalar intermediate states, see Fig.~\ref{fig:disp}$(a)$. The corresponding effects are determined by the respective TFFs, which can be inferred from experiment by dedicated dispersive analyses. Such a program for the $\pi^0$ pole~\cite{Schneider:2012ez,Hoferichter:2012pm,Hoferichter:2014vra,Hoferichter:2018dmo,Hoferichter:2018kwz,Hoferichter:2021lct} and the $\eta$, $\eta'$ poles~\cite{Stollenwerk:2011zz,Hanhart:2013vba,Kubis:2015sga,Holz:2015tcg,Holz:2022hwz,Holz:2022smu,Holz:2024lom,Holz:2024diw} led to the first entry in Table~\ref{tab:disp_summary}. Next, there are two-meson intermediate states, further decomposed into box contributions with two meson-pole left-hand cuts,
Fig.~\ref{fig:disp}$(b)$, and rescattering corrections, such as Fig.~\ref{fig:disp}$(c)$. The latter can be interpreted as a manifestation of the light scalar resonances, $f_0(500)$, $f_0(980)$, $a_0(980)$, in terms of helicity partial waves for $\gamma^*\gamma^*\to\pi\pi/\bar K K/\pi \eta$~\cite{Garcia-Martin:2010kyn,Hoferichter:2011wk,Moussallam:2013una,Hoferichter:2013ama,Danilkin:2018qfn,Hoferichter:2019nlq,Danilkin:2019opj,Lu:2020qeo,Schafer:2023qtl,Deineka:2024mzt}, see Table~\ref{tab:disp_summary} for a summary of the results.

\begin{table}[t]
	\centering
	\renewcommand{\arraystretch}{1.3}
	\begin{tabular}{lrr}
	\toprule
	 Contribution & $a_\mu [10^{-11}]$ & Reference\\\colrule
	 $\pi^0$, $\eta$, $\eta'$ poles & $91.2^{+2.9}_{-2.4}$ & \cite{Hoferichter:2018dmo,Hoferichter:2018kwz,Holz:2024lom,Holz:2024diw}\\
	 $\pi^\pm$ box & $-15.9(2)$ & \cite{Colangelo:2017qdm,Colangelo:2017fiz}\\
	 $K^\pm$ box & $-0.5(0)$ & \cite{Stamen:2022uqh}\\
	 $S$-wave rescattering & $-9.1(1.0)$ & \cite{Colangelo:2017qdm,Colangelo:2017fiz,Danilkin:2021icn,Deineka:2024mzt}\\\colrule
	 Sum & $65.7^{+3.1}_{-2.6}$ &\\
\botrule
	\renewcommand{\arraystretch}{1.0}
	\end{tabular}
	\caption{Exclusive hadronic states in the dispersive approach.}
	\label{tab:disp_summary}
\end{table}

Higher intermediate states, such as three-meson cuts in Fig.~\ref{fig:disp}$(d)$, are challenging to fully resolve, but by far the dominant effects arise from the resonant structures, e.g., the axial-vector states $a_1(1260)$ in the $3\pi$ channel, $f_1(1285)$ in $\pi\pi\eta$, and $f_1'(1420)$ in $\bar K K\pi$. Given the suppression by the respective masses, it is therefore sufficient to consider their contribution in a narrow-resonance approximation, as explicitly verified for $f_0(980)$ and $a_0(980)$ in Refs.~\cite{Danilkin:2021icn,Deineka:2024mzt}. For the corresponding evaluation of axial-vector states, we use the TFF parameterizations from Refs.~\cite{Zanke:2021wiq,Hoferichter:2023tgp}, relying on data for $e^+e^-\to e^+e^- A$, $A=f_1,f_1'$~\cite{Achard:2001uu,Achard:2007hm}, $e^+e^-\to f_1\pi^+\pi^-$~\cite{BaBar:2007qju,BaBar:2022ahi}, and radiative decays~\cite{Zanke:2021wiq,Hoferichter:2023tgp,ParticleDataGroup:2024cfk}. From the available data, the normalizations of the three TFFs for the $f_1$ and the mixing angle with the $f_1'$ can be extracted, which allows one to define a complete set of axial-vector TFFs using $U(3)$ symmetry. This assumption agrees well with a recent dispersive calculation of the $a_1$ singly-virtual form factor~\cite{Ludtke:2024ase}, which is also close to model predictions from holographic QCD~\cite{Leutgeb:2022lqw,Leutgeb:2024rfs}, but we assign a global $30\%$ uncertainty to the axial-vector contributions besides the uncertainties propagated from the fit parameters to account for possible symmetry violations. Finally, the TFF parameterizations need to be matched to the expected asymptotic behavior~\cite{Hoferichter:2020lap}, for which we developed a formulation that accounts for the effects of the axial-vector masses and leaves the low-energy properties of the TFFs unaltered, see Ref.~\cite{Hoferichter:2024bae} for details.

In addition to axial-vector states, also the tensor resonances $f_2(1270)$, $a_2(1320)$, and $f_2'(1525)$, as well as
the heavy scalars $f_0(1370)$ and $a_0(1450)$ could play a role. For the latter, we follow Ref.~\cite{Danilkin:2021icn}, identifying the scale in quark-model-inspired TFFs~\cite{Schuler:1997yw,Hoferichter:2020lap} with the mass of the $\rho(770)$, as supported by explicit calculations for $f_0(980)$ and $a_0(980)$~\cite{Danilkin:2021icn,Deineka:2024mzt}. For the tensor contributions, especially the $f_2(1270)$ as an elastic $\pi\pi$ resonance that arises predominantly from the unitarization of vector-meson left-hand cuts, it is also expected that the relevant scale is set by $\Mr$. Assuming the same quark-model-inspired form, all contributions but a single TFF vanish. Within this approximation, the problem of kinematic singularities is absent and the tensor contributions become amenable to an evaluation in the HLbL basis of Ref.~\cite{Hoferichter:2024fsj}. Accordingly, we include tensor estimates in this simplified set-up.

\begin{figure}[t]
     \centering
     \includegraphics[width=\linewidth]{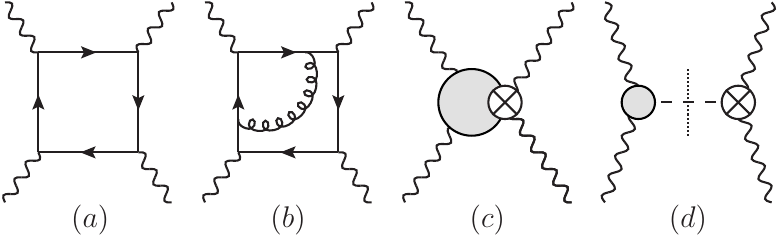}
     \caption{Diagrams relevant for SDCs: $(a)$ quark loop, $(b)$ $\alpha_s$ corrections, $(c)$ OPE, $(d)$ pseudoscalar pole in $VVA$. Solid lines refer to quarks, wiggly lines to photons, curly lines to gluons, and gray blobs to hadronic matrix elements. The cross indicates the OPE limit in which the two photons are replaced by an axial-vector current. }
     \label{fig:SDC}
 \end{figure}

 \emph{Matching to short-distance constraints}---The summation of exclusive states is  only feasible at sufficiently low energies, while beyond some scale $Q_0$ the representation needs to be matched to SDCs, see, e.g., Refs.~\cite{Leutgeb:2019gbz,Cappiello:2019hwh,Knecht:2020xyr,Masjuan:2020jsf,Ludtke:2020moa,Colangelo:2021nkr,Leutgeb:2024rfs,Eichmann:2024glq}. Our strategy for the matching proceeds as follows: in the region in which all $Q_i$ are larger than $Q_0$, we use the perturbative-QCD (pQCD) quark loop including $\alpha_s$ corrections, see Fig.~\ref{fig:SDC}$(a,b)$, using the $\alpha_s$ implementation from Refs.~\cite{Herren:2017osy,Chetyrkin:2000yt}. If all $Q_i$ are below $Q_0$, we use the hadronic description detailed above. In parts of the mixed region, e.g., $Q_3^2\ll (Q_1^2+Q_2^2)/2$, an OPE applies that relates the HLbL tensor in this limit to the $VVA$ correlator~\cite{Melnikov:2003xd,Vainshtein:2002nv,Knecht:2003xy}, see Fig.~\ref{fig:SDC}$(c)$. As input for the corresponding longitudinal and transverse form factors $w_{L,T}(q^2)$, we use the dispersive analysis from Ref.~\cite{Ludtke:2024ase}, including a generalization to singlet and octet components, see Fig.~\ref{fig:SDC}$(d)$ for the pseudoscalar-pole contributions and Ref.~\cite{Hoferichter:2024bae} for explicit representations of $w_{L,T}(q^2)$. The OPE constraint becomes particularly powerful due to a remarkable cancellation pointed out in Ref.~\cite{Bijnens:2024jgh}, which implies that at the level of the $a_\mu$ integration certain non-perturbative form factors that enter at higher orders in the OPE disappear. We use the OPE for
 \beq
Q_3^2\leq r \frac{Q_1^2+Q_2^2}{2},\quad Q_1^2\geq Q_0^2,\quad Q_2^2\geq Q_0^2,\quad Q_{3}^{2}\leq Q_{0}^{2},
\eeq
and similarly for small $Q_1^2$ or $Q_2^2$, with parameter $r$ varied within $[1/8,1/2]$. The matching scale $Q_0$ is varied between $1.2\GeV$, as the low scale at which the perturbative $\alpha_s$ corrections stay reasonably small, and $2.0\GeV$, as the high scale where the description in terms of the limited set of included hadronic states should still remain meaningful, with central values quoted for $Q_0=1.5\GeV$. In the OPE and pQCD regions we subtract the $\pi^0$, $\eta$, $\eta'$ poles to be able to continue to use the numbers from Table~\ref{tab:disp_summary}, while for the other contributions therein the overlap is negligible.

\begin{figure}[t]
     \centering
     \includegraphics[width=\linewidth]{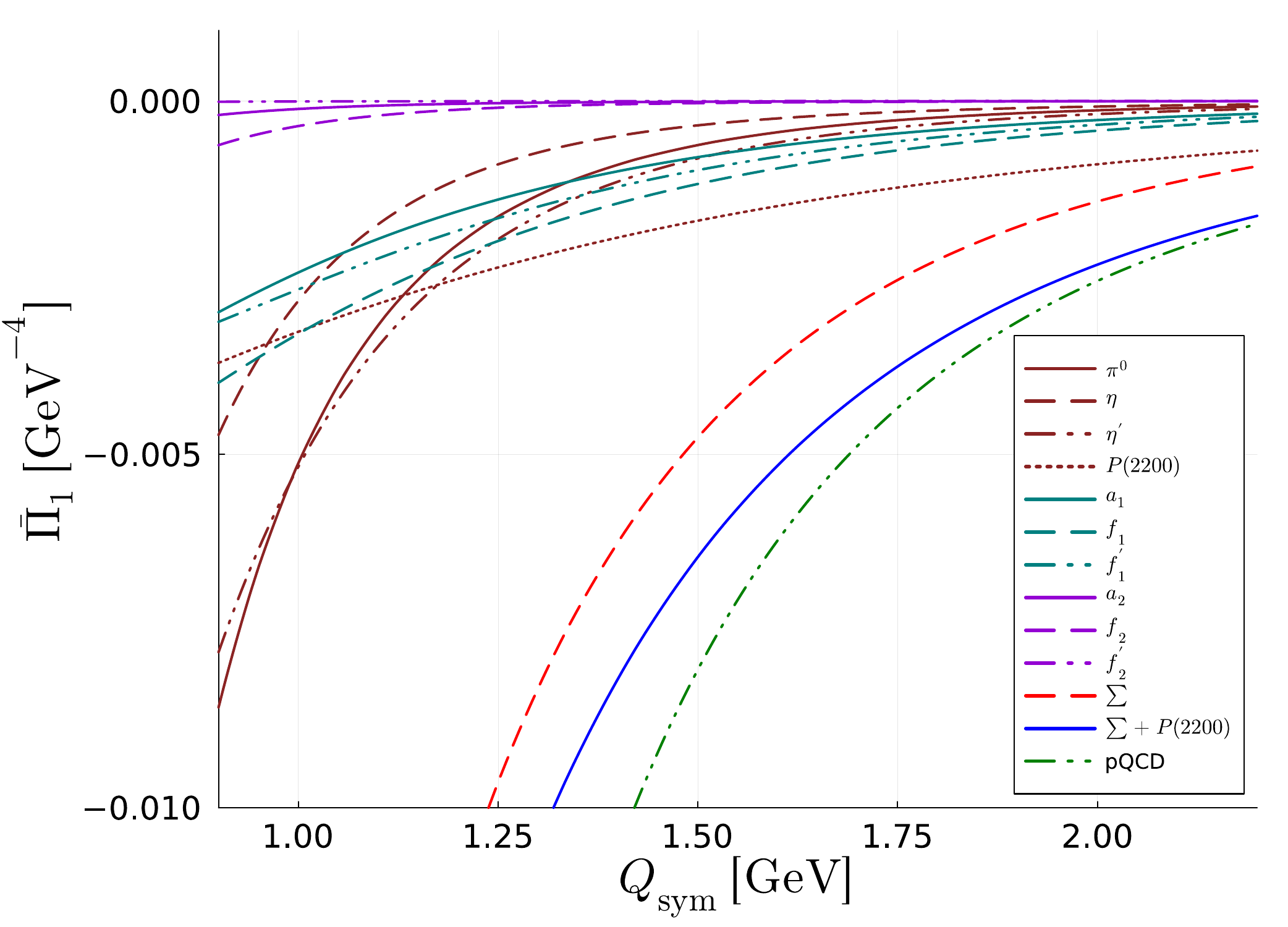}
     \caption{Matching between the sum of hadronic states, $\sum=\pi^0+\eta+\eta'+a_1+f_1+f_1'+a_2+f_2+f_2'$, in $\bar\Pi_1$ as a function of $Q_\text{sym}\equiv Q_1=Q_2=Q_3$. $P(2200)$ denotes the contribution of the effective pole, see main text.}
     \label{fig:matching}
 \end{figure}

 The matching between the sum of hadronic states and SDCs can be studied at the level of the scalar functions $\bar \Pi_i$, see Fig.~\ref{fig:matching} for an example. In general, the agreement is reasonable, especially in view of the limitations of either approach when extrapolated to the borders of the matching region, but we do observe some mismatch that could be interpreted as an effect of missing higher intermediate states. As a way to estimate the potential impact on the $a_\mu$ integral, we introduce an effective pole---pseudoscalar for $\bar\Pi_{1,2}$ and axial-vector for $\bar\Pi_{3\text{--}12}$, both in triangle kinematics~\cite{Ludtke:2023hvz} to be able to match the asymptotic behavior---with coefficients determined by the asymptotic matching, mass parameters $M_P^\text{eff}=2.2\GeV$ and $M_A^\text{eff}=1.7\GeV$, and TFF scale varied in the same range as $Q_0$. Figure~\ref{fig:matching} shows that imposing the exact asymptotic behavior indeed improves the matching in the intermediate region as well. Remarkably, the asymptotic coefficient comes out almost identical for all scalar functions that receive axial-vector contributions even when the matching is performed in the symmetric direction shown in Fig.~\ref{fig:matching}, leaving as dominant uncertainties the TFF scale and the difference to the asymmetric matching, when one virtuality is much smaller than the others. In particular, scanning over different choices for the determination of the effective-pole parameters, we assign an uncertainty that is larger than the entirety of the effective-pole contribution,
 see Ref.~\cite{Hoferichter:2024bae} for more details.

 \begin{table}[t]
	\centering
	\renewcommand{\arraystretch}{1.3}
	\begin{tabular}{llrrr}
	\toprule
	 Region & & $a_\mu[\bar\Pi_{1,2}] $ & $a_\mu[\bar\Pi_{3\text{--}12}] $ & Sum\\\colrule
\multirow{4}{*}{$Q_i<Q_0$} & $A$ &  $7.2(1.4)$ & $5.0(1.0)$ & $12.2(2.3)$\\
& $S$ & -- & $-0.7(3)$ & $-0.7(3)$\\
& $T$ & $2.6(3)$ & $-5.1(7)$ & $-2.5(3)$\\
& Eff.\ & $2.5$ & $-0.4$ & $2.0$\\\colrule
\multirow{3}{*}{Mixed} & $A,S,T$ & $2.5(7)$ & $1.3(3)$ & $3.8(1.0)$\\
& OPE & $6.3$ & $4.7$ & $10.9$\\
&Eff.\ & $1.1$ & $0.1$& $1.2$\\\colrule
$Q_i>Q_0$ & pQCD & $4.8^{+0.1}_{-0.2}$ & $1.6^{+0.0}_{-0.1}$ & $6.3^{+0.2}_{-0.3}$\\\colrule
Sum & & $26.9(2.1)$ & $6.3(1.5)$ & $33.2(3.3)$\\
\botrule
	\renewcommand{\arraystretch}{1.0}
	\end{tabular}
	\caption{Subleading contributions in the different integration regions, for $Q_0=1.5\GeV$, $r=1/4$, all values in units of $10^{-11}$. The hadronic states are $A=f_1,f_1',a_1$, $S=f_0(1370), a_0(1450)$, $T=f_2,a_2,f_2'$, ``Eff.'' refers to the effective poles ($P(2200)$ for $a_\mu[\bar\Pi_{1,2}]$ and $A(1700)$ for $a_\mu[\bar\Pi_{3\text{--}12}]$), with parameters determined from the matching in the symmetric asymptotic limit and TFF scale $1.5\GeV$. In the OPE and pQCD regions the $\pi^0$, $\eta$, $\eta'$ poles are subtracted.}
	\label{tab:subleading}
\end{table}

\begin{figure}[t]
     \centering
     \includegraphics[width=\linewidth]{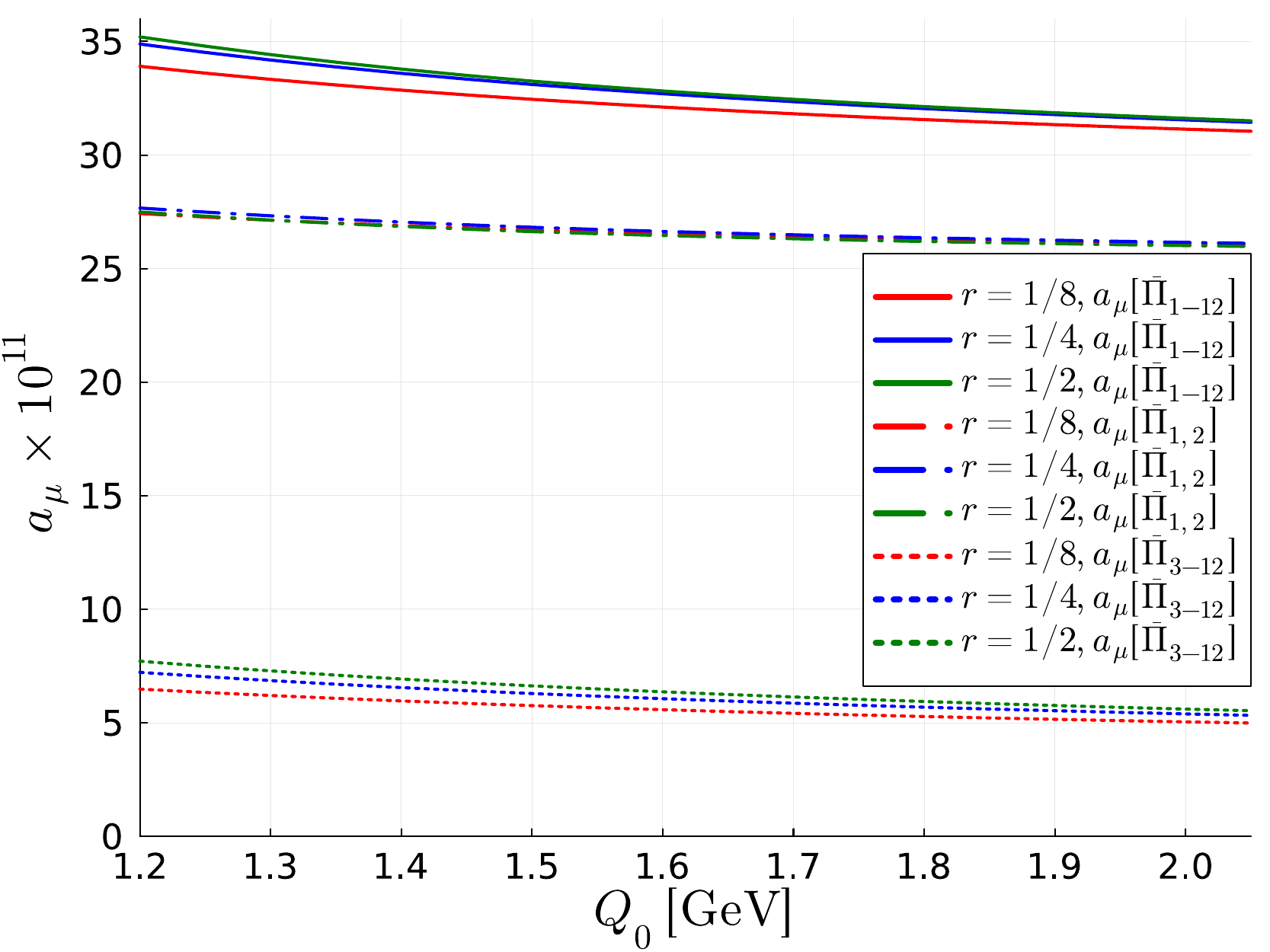}
     \caption{Stability of the $a_\mu$ integral under variation of $Q_0$, $r$.}
     \label{fig:Q0r}
 \end{figure}

\begin{figure}[t]
     \centering
     \includegraphics[width=\linewidth]{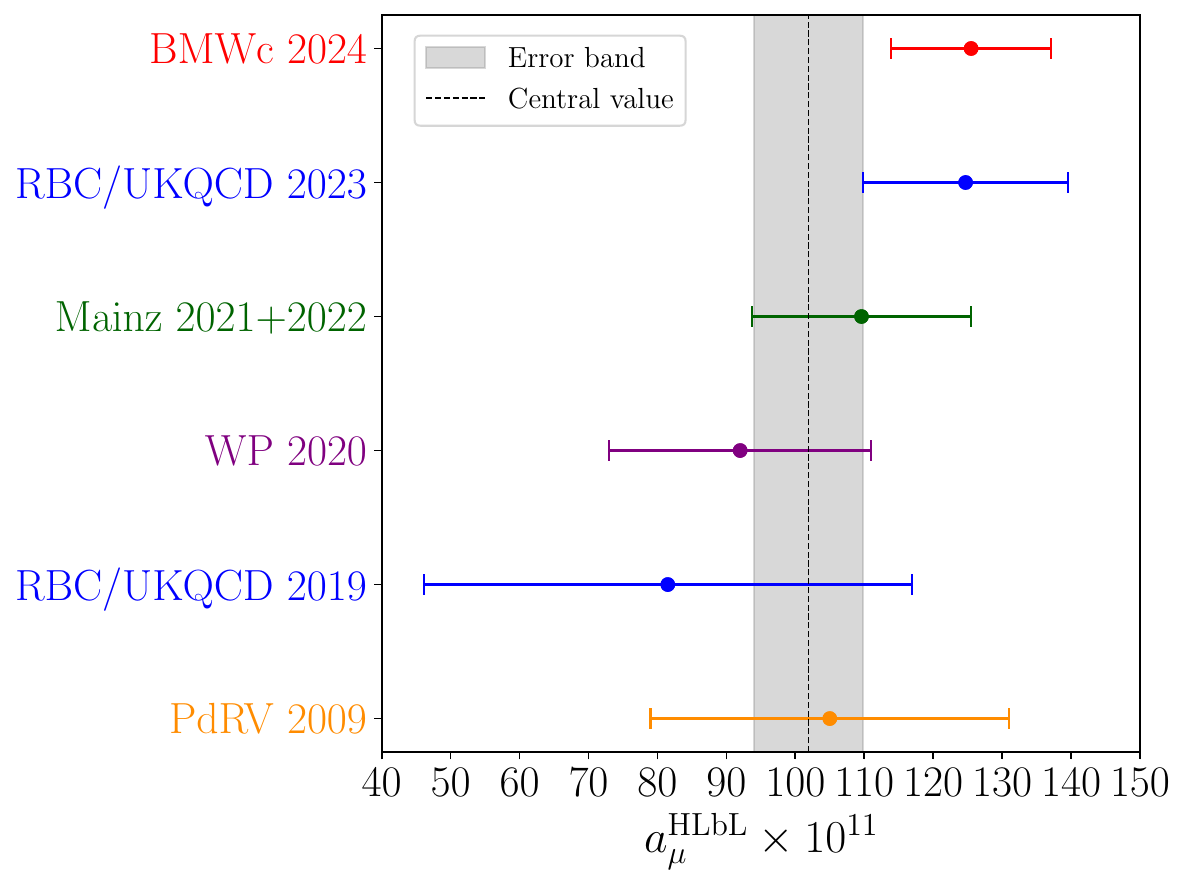}
     \caption{Comparison of our result for $a_\mu^\text{HLbL}$ (gray band) to the previous phenomenological evaluation from Ref.~\cite{Aoyama:2020ynm} (WP 2020) and the ``Glasgow consensus''~\cite{Prades:2009tw} (PdRV 2009), as well as the lattice-QCD calculations by RBC/UKQCD~\cite{Blum:2019ugy,Blum:2023vlm} (including the charm loop from Ref.~\cite{Chao:2022xzg}), Mainz~\cite{Chao:2021tvp,Chao:2022xzg},  and BMWc~\cite{Fodor:2024jyn}.}
     \label{fig:comparison}
 \end{figure}

\emph{Results for $a_\mu$}---The resulting values for the various subleading contributions are collected in Table~\ref{tab:subleading} for our central choice of $Q_0$ and $r$. The sensitivity to this choice is illustrated in Fig.~\ref{fig:Q0r}, demonstrating that the remaining matching uncertainty proves remarkably small.
In addition to (i) the experimental  (``exp'') errors already shown in Table~\ref{tab:subleading}, we include the following uncertainty estimates: (ii) matching (``match''), as the maximal variation for $Q_0\in[1.2,2.0]\GeV$, $r\in[1/8,1/2]$; (iii) systematic (``sys''), a $30\%$ uncertainty of the hadronic contributions to reflect the $U(3)$ assumptions for the axial-vector TFFs and the simplified tensor TFFs, plus an additional $100\%$ uncertainty on the total tensor contribution to protect against the cancellation observed between $a_\mu[\bar\Pi_{1,2}]$ and $a_\mu[\bar\Pi_{3\text{--}12}]$; (iv) effective pole (``eff''), reflecting the symmetric/asymmetric choices for the asymptotic matching and the sensitivity to the TFF scale. In this way, our final result for the subleading contributions becomes
 \begin{align}
\label{result_subleading}
 a_\mu[\bar\Pi_{1,2}]&=26.9(2.1)_\text{exp}(1.0)_\text{match}(3.7)_\text{sys}(3.2)_\text{eff}[5.4]_\text{total},\notag\\
 a_\mu[\bar\Pi_{3\text{--}12}]&=6.3(1.5)_\text{exp}(1.4)_\text{match}(0.2)_\text{sys}(2.2)_\text{eff}[3.0]_\text{total},\notag\\
 a_\mu[\bar\Pi_{1\text{--}12}]&=33.2(3.3)_\text{exp}(2.2)_\text{match}(4.6)_\text{sys}(3.9)_\text{eff}[7.2]_\text{total},
\end{align}
where all numbers are given in units of $10^{-11}$. Adding the dispersive numbers from Table~\ref{tab:disp_summary} and the charm loop, $a_\mu^{\text{HLbL}}[c]=3(1)\times 10^{-11}$~\cite{Colangelo:2019uex}, we obtain for the entire HLbL contribution
\beq
\label{amu_final}
a_\mu^\text{HLbL}=
101.9(7.9)\times 10^{-11}.
\eeq
This result agrees with the phenomenological evaluation from Ref.~\cite{Aoyama:2020ynm}, but reduces the uncertainty by more than a factor two.
Comparing to lattice QCD, our value is consistent with the QED$_L$ result by RBC/UKQCD~\cite{Blum:2019ugy} and the QED$_\infty$ one by the Mainz group~\cite{Chao:2021tvp,Chao:2022xzg}, while a small tension becomes visible compared to
the QED$_\infty$ evaluations by RBC/UKQCD~\cite{Blum:2023vlm} and BMWc~\cite{Fodor:2024jyn}, see Fig.~\ref{fig:comparison} for an overview.

Our final result~\eqref{amu_final} now meets the requirements defined by the projected precision of the Fermilab experiment. The substantial improvement derives from dedicated work on axial-vector contributions, SDCs, and the $VVA$ correlator, combined along the lines described in this Letter and detailed in Ref.~\cite{Hoferichter:2024bae}. In our opinion, this result
constitutes the best evaluation currently possible within a data-driven, dispersive approach, constraining all parts of the calculation to the furthest extent possible from data input and asymptotic matching.

Nevertheless, certain aspects of our uncertainty estimates~\eqref{result_subleading} should be corroborated and potentially improved in the future. First, the uncertainties in the axial-vector contributions should be validated with new data, which could become available at BESIII~\cite{Redmer:2024bva,BESIII:2020nme} and Belle II~\cite{Belle-II:2018jsg}. Second, the evaluation of the tensor contributions should be improved using a dispersive approach in triangle kinematics~\cite{Ludtke:2023hvz}, to assess explicitly the impact of the TFFs that vanish in the quark model or to even replace the narrow-width approximation for the $f_2(1270)$ by a dispersive treatment of $\pi\pi$-rescattering in the $D$-wave~\cite{Hoferichter:2019nlq,Danilkin:2019opj}. Third, developments along those lines, exploiting the complementarity between the two dispersive approaches, should lead to an improved matching to SDCs, and thereby reduce the sensitivity to effective-pole estimates as employed here. While these constitute important avenues for future improvements, we are convinced that Eq.~\eqref{amu_final} represents a realistic and conservative estimate of the current uncertainties.

 \emph{Acknowledgments}---We thank Gilberto Colangelo, Bastian Kubis, and Massimiliano Procura for decade-long collaboration on many aspects of the work presented here. We further thank
Johan Bijnens,  Nils Hermansson-Truedsson,  Jan L\"udtke, Antonio Rodr\'iguez-S\'anchez, and Matthias Steinhauser for valuable discussions, and Johan Bijnens for sharing code for the $\alpha_s$ corrections to the quark loop~\cite{Bijnens:2021jqo}.
Financial support by the SNSF (Project Nos.\ PCEFP2\_181117, PCEFP2\_194272, and TMCG-2\_213690) is gratefully acknowledged.

\vfill{}

\bibliography{amu}

\begin{thebibliography}{136}
\expandafter\ifx\csname natexlab\endcsname\relax\def\natexlab#1{#1}\fi
\expandafter\ifx\csname bibnamefont\endcsname\relax
  \def\bibnamefont#1{#1}\fi
\expandafter\ifx\csname bibfnamefont\endcsname\relax
  \def\bibfnamefont#1{#1}\fi
\expandafter\ifx\csname citenamefont\endcsname\relax
  \def\citenamefont#1{#1}\fi
\expandafter\ifx\csname url\endcsname\relax
  \def\url#1{\texttt{#1}}\fi
\expandafter\ifx\csname urlprefix\endcsname\relax\def\urlprefix{URL }\fi
\providecommand{\bibinfo}[2]{#2}
\providecommand{\eprint}[2][]{\url{#2}}

\bibitem[{\citenamefont{Schwinger}(1948)}]{Schwinger:1948iu}
\bibinfo{author}{\bibfnamefont{J.~S.} \bibnamefont{Schwinger}},
  \bibinfo{journal}{Phys. Rev.} \textbf{\bibinfo{volume}{73}},
  \bibinfo{pages}{416} (\bibinfo{year}{1948}).

\bibitem[{\citenamefont{Kusch and Foley}(1948)}]{Kusch:1948mvb}
\bibinfo{author}{\bibfnamefont{P.}~\bibnamefont{Kusch}} \bibnamefont{and}
  \bibinfo{author}{\bibfnamefont{H.~M.} \bibnamefont{Foley}},
  \bibinfo{journal}{Phys. Rev.} \textbf{\bibinfo{volume}{74}},
  \bibinfo{pages}{250} (\bibinfo{year}{1948}).

\bibitem[{\citenamefont{Aguillard et~al.}(2023)}]{Muong-2:2023cdq}
\bibinfo{author}{\bibfnamefont{D.~P.} \bibnamefont{Aguillard}}
  \bibnamefont{et~al.} (\bibinfo{collaboration}{Muon $g-2$}),
  \bibinfo{journal}{Phys. Rev. Lett.} \textbf{\bibinfo{volume}{131}},
  \bibinfo{pages}{161802} (\bibinfo{year}{2023}), \eprint{2308.06230}.

\bibitem[{\citenamefont{Aguillard et~al.}(2024)}]{Muong-2:2024hpx}
\bibinfo{author}{\bibfnamefont{D.~P.} \bibnamefont{Aguillard}}
  \bibnamefont{et~al.} (\bibinfo{collaboration}{Muon $g-2$}),
  \bibinfo{journal}{Phys. Rev. D} \textbf{\bibinfo{volume}{110}},
  \bibinfo{pages}{032009} (\bibinfo{year}{2024}), \eprint{2402.15410}.

\bibitem[{\citenamefont{Grange et~al.}(2015)}]{Muong-2:2015xgu}
\bibinfo{author}{\bibfnamefont{J.}~\bibnamefont{Grange}} \bibnamefont{et~al.}
  (\bibinfo{collaboration}{Muon $g-2$}) (\bibinfo{year}{2015}),
  \eprint{1501.06858}.

\bibitem[{\citenamefont{Aoyama et~al.}(2020)}]{Aoyama:2020ynm}
\bibinfo{author}{\bibfnamefont{T.}~\bibnamefont{Aoyama}} \bibnamefont{et~al.},
  \bibinfo{journal}{Phys. Rept.} \textbf{\bibinfo{volume}{887}},
  \bibinfo{pages}{1} (\bibinfo{year}{2020}), \eprint{2006.04822}.

\bibitem[{\citenamefont{Aoyama et~al.}(2012)\citenamefont{Aoyama, Hayakawa,
  Kinoshita, and Nio}}]{Aoyama:2012wk}
\bibinfo{author}{\bibfnamefont{T.}~\bibnamefont{Aoyama}},
  \bibinfo{author}{\bibfnamefont{M.}~\bibnamefont{Hayakawa}},
  \bibinfo{author}{\bibfnamefont{T.}~\bibnamefont{Kinoshita}},
  \bibnamefont{and} \bibinfo{author}{\bibfnamefont{M.}~\bibnamefont{Nio}},
  \bibinfo{journal}{Phys. Rev. Lett.} \textbf{\bibinfo{volume}{109}},
  \bibinfo{pages}{111808} (\bibinfo{year}{2012}), \eprint{1205.5370}.

\bibitem[{\citenamefont{Aoyama et~al.}(2019)\citenamefont{Aoyama, Kinoshita,
  and Nio}}]{Aoyama:2019ryr}
\bibinfo{author}{\bibfnamefont{T.}~\bibnamefont{Aoyama}},
  \bibinfo{author}{\bibfnamefont{T.}~\bibnamefont{Kinoshita}},
  \bibnamefont{and} \bibinfo{author}{\bibfnamefont{M.}~\bibnamefont{Nio}},
  \bibinfo{journal}{Atoms} \textbf{\bibinfo{volume}{7}}, \bibinfo{pages}{28}
  (\bibinfo{year}{2019}).

\bibitem[{\citenamefont{Czarnecki et~al.}(2003)\citenamefont{Czarnecki,
  Marciano, and Vainshtein}}]{Czarnecki:2002nt}
\bibinfo{author}{\bibfnamefont{A.}~\bibnamefont{Czarnecki}},
  \bibinfo{author}{\bibfnamefont{W.~J.} \bibnamefont{Marciano}},
  \bibnamefont{and}
  \bibinfo{author}{\bibfnamefont{A.}~\bibnamefont{Vainshtein}},
  \bibinfo{journal}{Phys. Rev. D} \textbf{\bibinfo{volume}{67}},
  \bibinfo{pages}{073006} (\bibinfo{year}{2003}), \bibinfo{note}{[Erratum:
  Phys. Rev. D {\bf 73}, 119901 (2006)]}, \eprint{hep-ph/0212229}.

\bibitem[{\citenamefont{Gnendiger et~al.}(2013)\citenamefont{Gnendiger,
  St\"ockinger, and St\"ockinger-Kim}}]{Gnendiger:2013pva}
\bibinfo{author}{\bibfnamefont{C.}~\bibnamefont{Gnendiger}},
  \bibinfo{author}{\bibfnamefont{D.}~\bibnamefont{St\"ockinger}},
  \bibnamefont{and}
  \bibinfo{author}{\bibfnamefont{H.}~\bibnamefont{St\"ockinger-Kim}},
  \bibinfo{journal}{Phys. Rev. D} \textbf{\bibinfo{volume}{88}},
  \bibinfo{pages}{053005} (\bibinfo{year}{2013}), \eprint{1306.5546}.

\bibitem[{\citenamefont{Davier et~al.}(2017)\citenamefont{Davier, Hoecker,
  Malaescu, and Zhang}}]{Davier:2017zfy}
\bibinfo{author}{\bibfnamefont{M.}~\bibnamefont{Davier}},
  \bibinfo{author}{\bibfnamefont{A.}~\bibnamefont{Hoecker}},
  \bibinfo{author}{\bibfnamefont{B.}~\bibnamefont{Malaescu}}, \bibnamefont{and}
  \bibinfo{author}{\bibfnamefont{Z.}~\bibnamefont{Zhang}},
  \bibinfo{journal}{Eur. Phys. J. C} \textbf{\bibinfo{volume}{77}},
  \bibinfo{pages}{827} (\bibinfo{year}{2017}), \eprint{1706.09436}.

\bibitem[{\citenamefont{Keshavarzi et~al.}(2018)\citenamefont{Keshavarzi,
  Nomura, and Teubner}}]{Keshavarzi:2018mgv}
\bibinfo{author}{\bibfnamefont{A.}~\bibnamefont{Keshavarzi}},
  \bibinfo{author}{\bibfnamefont{D.}~\bibnamefont{Nomura}}, \bibnamefont{and}
  \bibinfo{author}{\bibfnamefont{T.}~\bibnamefont{Teubner}},
  \bibinfo{journal}{Phys. Rev. D} \textbf{\bibinfo{volume}{97}},
  \bibinfo{pages}{114025} (\bibinfo{year}{2018}), \eprint{1802.02995}.

\bibitem[{\citenamefont{Colangelo et~al.}(2019)\citenamefont{Colangelo,
  Hoferichter, and Stoffer}}]{Colangelo:2018mtw}
\bibinfo{author}{\bibfnamefont{G.}~\bibnamefont{Colangelo}},
  \bibinfo{author}{\bibfnamefont{M.}~\bibnamefont{Hoferichter}},
  \bibnamefont{and} \bibinfo{author}{\bibfnamefont{P.}~\bibnamefont{Stoffer}},
  \bibinfo{journal}{JHEP} \textbf{\bibinfo{volume}{02}}, \bibinfo{pages}{006}
  (\bibinfo{year}{2019}), \eprint{1810.00007}.

\bibitem[{\citenamefont{Hoferichter et~al.}(2019)\citenamefont{Hoferichter,
  Hoid, and Kubis}}]{Hoferichter:2019gzf}
\bibinfo{author}{\bibfnamefont{M.}~\bibnamefont{Hoferichter}},
  \bibinfo{author}{\bibfnamefont{B.-L.} \bibnamefont{Hoid}}, \bibnamefont{and}
  \bibinfo{author}{\bibfnamefont{B.}~\bibnamefont{Kubis}},
  \bibinfo{journal}{JHEP} \textbf{\bibinfo{volume}{08}}, \bibinfo{pages}{137}
  (\bibinfo{year}{2019}), \eprint{1907.01556}.

\bibitem[{\citenamefont{Davier et~al.}(2020)\citenamefont{Davier, Hoecker,
  Malaescu, and Zhang}}]{Davier:2019can}
\bibinfo{author}{\bibfnamefont{M.}~\bibnamefont{Davier}},
  \bibinfo{author}{\bibfnamefont{A.}~\bibnamefont{Hoecker}},
  \bibinfo{author}{\bibfnamefont{B.}~\bibnamefont{Malaescu}}, \bibnamefont{and}
  \bibinfo{author}{\bibfnamefont{Z.}~\bibnamefont{Zhang}},
  \bibinfo{journal}{Eur. Phys. J. C} \textbf{\bibinfo{volume}{80}},
  \bibinfo{pages}{241} (\bibinfo{year}{2020}), \bibinfo{note}{[Erratum: Eur.
  Phys. J. C {\bf 80}, 410 (2020)]}, \eprint{1908.00921}.

\bibitem[{\citenamefont{Keshavarzi
  et~al.}(2020{\natexlab{a}})\citenamefont{Keshavarzi, Nomura, and
  Teubner}}]{Keshavarzi:2019abf}
\bibinfo{author}{\bibfnamefont{A.}~\bibnamefont{Keshavarzi}},
  \bibinfo{author}{\bibfnamefont{D.}~\bibnamefont{Nomura}}, \bibnamefont{and}
  \bibinfo{author}{\bibfnamefont{T.}~\bibnamefont{Teubner}},
  \bibinfo{journal}{Phys. Rev. D} \textbf{\bibinfo{volume}{101}},
  \bibinfo{pages}{014029} (\bibinfo{year}{2020}{\natexlab{a}}),
  \eprint{1911.00367}.

\bibitem[{\citenamefont{Hoid et~al.}(2020)\citenamefont{Hoid, Hoferichter, and
  Kubis}}]{Hoid:2020xjs}
\bibinfo{author}{\bibfnamefont{B.-L.} \bibnamefont{Hoid}},
  \bibinfo{author}{\bibfnamefont{M.}~\bibnamefont{Hoferichter}},
  \bibnamefont{and} \bibinfo{author}{\bibfnamefont{B.}~\bibnamefont{Kubis}},
  \bibinfo{journal}{Eur. Phys. J. C} \textbf{\bibinfo{volume}{80}},
  \bibinfo{pages}{988} (\bibinfo{year}{2020}), \eprint{2007.12696}.

\bibitem[{\citenamefont{Melnikov and Vainshtein}(2004)}]{Melnikov:2003xd}
\bibinfo{author}{\bibfnamefont{K.}~\bibnamefont{Melnikov}} \bibnamefont{and}
  \bibinfo{author}{\bibfnamefont{A.}~\bibnamefont{Vainshtein}},
  \bibinfo{journal}{Phys. Rev. D} \textbf{\bibinfo{volume}{70}},
  \bibinfo{pages}{113006} (\bibinfo{year}{2004}), \eprint{hep-ph/0312226}.

\bibitem[{\citenamefont{Masjuan and
  S{\'a}nchez-Puertas}(2017)}]{Masjuan:2017tvw}
\bibinfo{author}{\bibfnamefont{P.}~\bibnamefont{Masjuan}} \bibnamefont{and}
  \bibinfo{author}{\bibfnamefont{P.}~\bibnamefont{S{\'a}nchez-Puertas}},
  \bibinfo{journal}{Phys. Rev. D} \textbf{\bibinfo{volume}{95}},
  \bibinfo{pages}{054026} (\bibinfo{year}{2017}), \eprint{1701.05829}.

\bibitem[{\citenamefont{Colangelo
  et~al.}(2017{\natexlab{a}})\citenamefont{Colangelo, Hoferichter, Procura, and
  Stoffer}}]{Colangelo:2017qdm}
\bibinfo{author}{\bibfnamefont{G.}~\bibnamefont{Colangelo}},
  \bibinfo{author}{\bibfnamefont{M.}~\bibnamefont{Hoferichter}},
  \bibinfo{author}{\bibfnamefont{M.}~\bibnamefont{Procura}}, \bibnamefont{and}
  \bibinfo{author}{\bibfnamefont{P.}~\bibnamefont{Stoffer}},
  \bibinfo{journal}{Phys. Rev. Lett.} \textbf{\bibinfo{volume}{118}},
  \bibinfo{pages}{232001} (\bibinfo{year}{2017}{\natexlab{a}}),
  \eprint{1701.06554}.

\bibitem[{\citenamefont{Colangelo
  et~al.}(2017{\natexlab{b}})\citenamefont{Colangelo, Hoferichter, Procura, and
  Stoffer}}]{Colangelo:2017fiz}
\bibinfo{author}{\bibfnamefont{G.}~\bibnamefont{Colangelo}},
  \bibinfo{author}{\bibfnamefont{M.}~\bibnamefont{Hoferichter}},
  \bibinfo{author}{\bibfnamefont{M.}~\bibnamefont{Procura}}, \bibnamefont{and}
  \bibinfo{author}{\bibfnamefont{P.}~\bibnamefont{Stoffer}},
  \bibinfo{journal}{JHEP} \textbf{\bibinfo{volume}{04}}, \bibinfo{pages}{161}
  (\bibinfo{year}{2017}{\natexlab{b}}), \eprint{1702.07347}.

\bibitem[{\citenamefont{Hoferichter
  et~al.}(2018{\natexlab{a}})\citenamefont{Hoferichter, Hoid, Kubis, Leupold,
  and Schneider}}]{Hoferichter:2018dmo}
\bibinfo{author}{\bibfnamefont{M.}~\bibnamefont{Hoferichter}},
  \bibinfo{author}{\bibfnamefont{B.-L.} \bibnamefont{Hoid}},
  \bibinfo{author}{\bibfnamefont{B.}~\bibnamefont{Kubis}},
  \bibinfo{author}{\bibfnamefont{S.}~\bibnamefont{Leupold}}, \bibnamefont{and}
  \bibinfo{author}{\bibfnamefont{S.~P.} \bibnamefont{Schneider}},
  \bibinfo{journal}{Phys. Rev. Lett.} \textbf{\bibinfo{volume}{121}},
  \bibinfo{pages}{112002} (\bibinfo{year}{2018}{\natexlab{a}}),
  \eprint{1805.01471}.

\bibitem[{\citenamefont{Hoferichter
  et~al.}(2018{\natexlab{b}})\citenamefont{Hoferichter, Hoid, Kubis, Leupold,
  and Schneider}}]{Hoferichter:2018kwz}
\bibinfo{author}{\bibfnamefont{M.}~\bibnamefont{Hoferichter}},
  \bibinfo{author}{\bibfnamefont{B.-L.} \bibnamefont{Hoid}},
  \bibinfo{author}{\bibfnamefont{B.}~\bibnamefont{Kubis}},
  \bibinfo{author}{\bibfnamefont{S.}~\bibnamefont{Leupold}}, \bibnamefont{and}
  \bibinfo{author}{\bibfnamefont{S.~P.} \bibnamefont{Schneider}},
  \bibinfo{journal}{JHEP} \textbf{\bibinfo{volume}{10}}, \bibinfo{pages}{141}
  (\bibinfo{year}{2018}{\natexlab{b}}), \eprint{1808.04823}.

\bibitem[{\citenamefont{G\'erardin et~al.}(2019)\citenamefont{G\'erardin,
  Meyer, and Nyffeler}}]{Gerardin:2019vio}
\bibinfo{author}{\bibfnamefont{A.}~\bibnamefont{G\'erardin}},
  \bibinfo{author}{\bibfnamefont{H.~B.} \bibnamefont{Meyer}}, \bibnamefont{and}
  \bibinfo{author}{\bibfnamefont{A.}~\bibnamefont{Nyffeler}},
  \bibinfo{journal}{Phys. Rev. D} \textbf{\bibinfo{volume}{100}},
  \bibinfo{pages}{034520} (\bibinfo{year}{2019}), \eprint{1903.09471}.

\bibitem[{\citenamefont{Bijnens et~al.}(2019)\citenamefont{Bijnens,
  Hermansson-Truedsson, and Rodr\'\i{}guez-S\'anchez}}]{Bijnens:2019ghy}
\bibinfo{author}{\bibfnamefont{J.}~\bibnamefont{Bijnens}},
  \bibinfo{author}{\bibfnamefont{N.}~\bibnamefont{Hermansson-Truedsson}},
  \bibnamefont{and}
  \bibinfo{author}{\bibfnamefont{A.}~\bibnamefont{Rodr\'\i{}guez-S\'anchez}},
  \bibinfo{journal}{Phys. Lett. B} \textbf{\bibinfo{volume}{798}},
  \bibinfo{pages}{134994} (\bibinfo{year}{2019}), \eprint{1908.03331}.

\bibitem[{\citenamefont{Colangelo
  et~al.}(2020{\natexlab{a}})\citenamefont{Colangelo, Hagelstein, Hoferichter,
  Laub, and Stoffer}}]{Colangelo:2019lpu}
\bibinfo{author}{\bibfnamefont{G.}~\bibnamefont{Colangelo}},
  \bibinfo{author}{\bibfnamefont{F.}~\bibnamefont{Hagelstein}},
  \bibinfo{author}{\bibfnamefont{M.}~\bibnamefont{Hoferichter}},
  \bibinfo{author}{\bibfnamefont{L.}~\bibnamefont{Laub}}, \bibnamefont{and}
  \bibinfo{author}{\bibfnamefont{P.}~\bibnamefont{Stoffer}},
  \bibinfo{journal}{Phys. Rev. D} \textbf{\bibinfo{volume}{101}},
  \bibinfo{pages}{051501} (\bibinfo{year}{2020}{\natexlab{a}}),
  \eprint{1910.11881}.

\bibitem[{\citenamefont{Colangelo
  et~al.}(2020{\natexlab{b}})\citenamefont{Colangelo, Hagelstein, Hoferichter,
  Laub, and Stoffer}}]{Colangelo:2019uex}
\bibinfo{author}{\bibfnamefont{G.}~\bibnamefont{Colangelo}},
  \bibinfo{author}{\bibfnamefont{F.}~\bibnamefont{Hagelstein}},
  \bibinfo{author}{\bibfnamefont{M.}~\bibnamefont{Hoferichter}},
  \bibinfo{author}{\bibfnamefont{L.}~\bibnamefont{Laub}}, \bibnamefont{and}
  \bibinfo{author}{\bibfnamefont{P.}~\bibnamefont{Stoffer}},
  \bibinfo{journal}{JHEP} \textbf{\bibinfo{volume}{03}}, \bibinfo{pages}{101}
  (\bibinfo{year}{2020}{\natexlab{b}}), \eprint{1910.13432}.

\bibitem[{\citenamefont{Pauk and Vanderhaeghen}(2014)}]{Pauk:2014rta}
\bibinfo{author}{\bibfnamefont{V.}~\bibnamefont{Pauk}} \bibnamefont{and}
  \bibinfo{author}{\bibfnamefont{M.}~\bibnamefont{Vanderhaeghen}},
  \bibinfo{journal}{Eur. Phys. J. C} \textbf{\bibinfo{volume}{74}},
  \bibinfo{pages}{3008} (\bibinfo{year}{2014}), \eprint{1401.0832}.

\bibitem[{\citenamefont{Danilkin and Vanderhaeghen}(2017)}]{Danilkin:2016hnh}
\bibinfo{author}{\bibfnamefont{I.}~\bibnamefont{Danilkin}} \bibnamefont{and}
  \bibinfo{author}{\bibfnamefont{M.}~\bibnamefont{Vanderhaeghen}},
  \bibinfo{journal}{Phys. Rev. D} \textbf{\bibinfo{volume}{95}},
  \bibinfo{pages}{014019} (\bibinfo{year}{2017}), \eprint{1611.04646}.

\bibitem[{\citenamefont{Jegerlehner}(2017)}]{Jegerlehner:2017gek}
\bibinfo{author}{\bibfnamefont{F.}~\bibnamefont{Jegerlehner}},
  \emph{\bibinfo{title}{{The Anomalous Magnetic Moment of the Muon}}}, vol.
  \bibinfo{volume}{274} (\bibinfo{publisher}{Springer},
  \bibinfo{address}{Cham}, \bibinfo{year}{2017}).

\bibitem[{\citenamefont{Knecht et~al.}(2018)\citenamefont{Knecht, Narison,
  Rabemananjara, and Rabetiarivony}}]{Knecht:2018sci}
\bibinfo{author}{\bibfnamefont{M.}~\bibnamefont{Knecht}},
  \bibinfo{author}{\bibfnamefont{S.}~\bibnamefont{Narison}},
  \bibinfo{author}{\bibfnamefont{A.}~\bibnamefont{Rabemananjara}},
  \bibnamefont{and}
  \bibinfo{author}{\bibfnamefont{D.}~\bibnamefont{Rabetiarivony}},
  \bibinfo{journal}{Phys. Lett. B} \textbf{\bibinfo{volume}{787}},
  \bibinfo{pages}{111} (\bibinfo{year}{2018}), \eprint{1808.03848}.

\bibitem[{\citenamefont{Eichmann et~al.}(2020)\citenamefont{Eichmann, Fischer,
  and Williams}}]{Eichmann:2019bqf}
\bibinfo{author}{\bibfnamefont{G.}~\bibnamefont{Eichmann}},
  \bibinfo{author}{\bibfnamefont{C.~S.} \bibnamefont{Fischer}},
  \bibnamefont{and} \bibinfo{author}{\bibfnamefont{R.}~\bibnamefont{Williams}},
  \bibinfo{journal}{Phys. Rev. D} \textbf{\bibinfo{volume}{101}},
  \bibinfo{pages}{054015} (\bibinfo{year}{2020}), \eprint{1910.06795}.

\bibitem[{\citenamefont{Roig and S{\'a}nchez-Puertas}(2020)}]{Roig:2019reh}
\bibinfo{author}{\bibfnamefont{P.}~\bibnamefont{Roig}} \bibnamefont{and}
  \bibinfo{author}{\bibfnamefont{P.}~\bibnamefont{S{\'a}nchez-Puertas}},
  \bibinfo{journal}{Phys. Rev. D} \textbf{\bibinfo{volume}{101}},
  \bibinfo{pages}{074019} (\bibinfo{year}{2020}), \eprint{1910.02881}.

\bibitem[{\citenamefont{Calmet et~al.}(1976)\citenamefont{Calmet, Narison,
  Perrottet, and de~Rafael}}]{Calmet:1976kd}
\bibinfo{author}{\bibfnamefont{J.}~\bibnamefont{Calmet}},
  \bibinfo{author}{\bibfnamefont{S.}~\bibnamefont{Narison}},
  \bibinfo{author}{\bibfnamefont{M.}~\bibnamefont{Perrottet}},
  \bibnamefont{and}
  \bibinfo{author}{\bibfnamefont{E.}~\bibnamefont{de~Rafael}},
  \bibinfo{journal}{Phys. Lett. B} \textbf{\bibinfo{volume}{61}},
  \bibinfo{pages}{283} (\bibinfo{year}{1976}).

\bibitem[{\citenamefont{Kurz et~al.}(2014)\citenamefont{Kurz, Liu, Marquard,
  and Steinhauser}}]{Kurz:2014wya}
\bibinfo{author}{\bibfnamefont{A.}~\bibnamefont{Kurz}},
  \bibinfo{author}{\bibfnamefont{T.}~\bibnamefont{Liu}},
  \bibinfo{author}{\bibfnamefont{P.}~\bibnamefont{Marquard}}, \bibnamefont{and}
  \bibinfo{author}{\bibfnamefont{M.}~\bibnamefont{Steinhauser}},
  \bibinfo{journal}{Phys. Lett. B} \textbf{\bibinfo{volume}{734}},
  \bibinfo{pages}{144} (\bibinfo{year}{2014}), \eprint{1403.6400}.

\bibitem[{\citenamefont{Colangelo
  et~al.}(2014{\natexlab{a}})\citenamefont{Colangelo, Hoferichter, Nyffeler,
  Passera, and Stoffer}}]{Colangelo:2014qya}
\bibinfo{author}{\bibfnamefont{G.}~\bibnamefont{Colangelo}},
  \bibinfo{author}{\bibfnamefont{M.}~\bibnamefont{Hoferichter}},
  \bibinfo{author}{\bibfnamefont{A.}~\bibnamefont{Nyffeler}},
  \bibinfo{author}{\bibfnamefont{M.}~\bibnamefont{Passera}}, \bibnamefont{and}
  \bibinfo{author}{\bibfnamefont{P.}~\bibnamefont{Stoffer}},
  \bibinfo{journal}{Phys. Lett. B} \textbf{\bibinfo{volume}{735}},
  \bibinfo{pages}{90} (\bibinfo{year}{2014}{\natexlab{a}}), \eprint{1403.7512}.

\bibitem[{\citenamefont{Hoferichter and Teubner}(2022)}]{Hoferichter:2021wyj}
\bibinfo{author}{\bibfnamefont{M.}~\bibnamefont{Hoferichter}} \bibnamefont{and}
  \bibinfo{author}{\bibfnamefont{T.}~\bibnamefont{Teubner}},
  \bibinfo{journal}{Phys. Rev. Lett.} \textbf{\bibinfo{volume}{128}},
  \bibinfo{pages}{112002} (\bibinfo{year}{2022}), \eprint{2112.06929}.

\bibitem[{\citenamefont{Crivellin et~al.}(2020)\citenamefont{Crivellin,
  Hoferichter, Manzari, and Montull}}]{Crivellin:2020zul}
\bibinfo{author}{\bibfnamefont{A.}~\bibnamefont{Crivellin}},
  \bibinfo{author}{\bibfnamefont{M.}~\bibnamefont{Hoferichter}},
  \bibinfo{author}{\bibfnamefont{C.~A.} \bibnamefont{Manzari}},
  \bibnamefont{and} \bibinfo{author}{\bibfnamefont{M.}~\bibnamefont{Montull}},
  \bibinfo{journal}{Phys. Rev. Lett.} \textbf{\bibinfo{volume}{125}},
  \bibinfo{pages}{091801} (\bibinfo{year}{2020}), \eprint{2003.04886}.

\bibitem[{\citenamefont{Keshavarzi
  et~al.}(2020{\natexlab{b}})\citenamefont{Keshavarzi, Marciano, Passera, and
  Sirlin}}]{Keshavarzi:2020bfy}
\bibinfo{author}{\bibfnamefont{A.}~\bibnamefont{Keshavarzi}},
  \bibinfo{author}{\bibfnamefont{W.~J.} \bibnamefont{Marciano}},
  \bibinfo{author}{\bibfnamefont{M.}~\bibnamefont{Passera}}, \bibnamefont{and}
  \bibinfo{author}{\bibfnamefont{A.}~\bibnamefont{Sirlin}},
  \bibinfo{journal}{Phys. Rev. D} \textbf{\bibinfo{volume}{102}},
  \bibinfo{pages}{033002} (\bibinfo{year}{2020}{\natexlab{b}}),
  \eprint{2006.12666}.

\bibitem[{\citenamefont{Malaescu and Schott}(2021)}]{Malaescu:2020zuc}
\bibinfo{author}{\bibfnamefont{B.}~\bibnamefont{Malaescu}} \bibnamefont{and}
  \bibinfo{author}{\bibfnamefont{M.}~\bibnamefont{Schott}},
  \bibinfo{journal}{Eur. Phys. J. C} \textbf{\bibinfo{volume}{81}},
  \bibinfo{pages}{46} (\bibinfo{year}{2021}), \eprint{2008.08107}.

\bibitem[{\citenamefont{Colangelo
  et~al.}(2021{\natexlab{a}})\citenamefont{Colangelo, Hoferichter, and
  Stoffer}}]{Colangelo:2020lcg}
\bibinfo{author}{\bibfnamefont{G.}~\bibnamefont{Colangelo}},
  \bibinfo{author}{\bibfnamefont{M.}~\bibnamefont{Hoferichter}},
  \bibnamefont{and} \bibinfo{author}{\bibfnamefont{P.}~\bibnamefont{Stoffer}},
  \bibinfo{journal}{Phys. Lett. B} \textbf{\bibinfo{volume}{814}},
  \bibinfo{pages}{136073} (\bibinfo{year}{2021}{\natexlab{a}}),
  \eprint{2010.07943}.

\bibitem[{\citenamefont{Stamen et~al.}(2022)\citenamefont{Stamen, Hariharan,
  Hoferichter, Kubis, and Stoffer}}]{Stamen:2022uqh}
\bibinfo{author}{\bibfnamefont{D.}~\bibnamefont{Stamen}},
  \bibinfo{author}{\bibfnamefont{D.}~\bibnamefont{Hariharan}},
  \bibinfo{author}{\bibfnamefont{M.}~\bibnamefont{Hoferichter}},
  \bibinfo{author}{\bibfnamefont{B.}~\bibnamefont{Kubis}}, \bibnamefont{and}
  \bibinfo{author}{\bibfnamefont{P.}~\bibnamefont{Stoffer}},
  \bibinfo{journal}{Eur. Phys. J. C} \textbf{\bibinfo{volume}{82}},
  \bibinfo{pages}{432} (\bibinfo{year}{2022}), \eprint{2202.11106}.

\bibitem[{\citenamefont{Colangelo
  et~al.}(2022{\natexlab{a}})\citenamefont{Colangelo, El-Khadra, Hoferichter,
  Keshavarzi, Lehner, Stoffer, and Teubner}}]{Colangelo:2022vok}
\bibinfo{author}{\bibfnamefont{G.}~\bibnamefont{Colangelo}},
  \bibinfo{author}{\bibfnamefont{A.~X.} \bibnamefont{El-Khadra}},
  \bibinfo{author}{\bibfnamefont{M.}~\bibnamefont{Hoferichter}},
  \bibinfo{author}{\bibfnamefont{A.}~\bibnamefont{Keshavarzi}},
  \bibinfo{author}{\bibfnamefont{C.}~\bibnamefont{Lehner}},
  \bibinfo{author}{\bibfnamefont{P.}~\bibnamefont{Stoffer}}, \bibnamefont{and}
  \bibinfo{author}{\bibfnamefont{T.}~\bibnamefont{Teubner}},
  \bibinfo{journal}{Phys. Lett. B} \textbf{\bibinfo{volume}{833}},
  \bibinfo{pages}{137313} (\bibinfo{year}{2022}{\natexlab{a}}),
  \eprint{2205.12963}.

\bibitem[{\citenamefont{Colangelo
  et~al.}(2022{\natexlab{b}})\citenamefont{Colangelo, Hoferichter, Kubis, and
  Stoffer}}]{Colangelo:2022prz}
\bibinfo{author}{\bibfnamefont{G.}~\bibnamefont{Colangelo}},
  \bibinfo{author}{\bibfnamefont{M.}~\bibnamefont{Hoferichter}},
  \bibinfo{author}{\bibfnamefont{B.}~\bibnamefont{Kubis}}, \bibnamefont{and}
  \bibinfo{author}{\bibfnamefont{P.}~\bibnamefont{Stoffer}},
  \bibinfo{journal}{JHEP} \textbf{\bibinfo{volume}{10}}, \bibinfo{pages}{032}
  (\bibinfo{year}{2022}{\natexlab{b}}), \eprint{2208.08993}.

\bibitem[{\citenamefont{Hoferichter
  et~al.}(2023{\natexlab{a}})\citenamefont{Hoferichter, Colangelo, Hoid, Kubis,
  Ruiz~de Elvira, Schuh, Stamen, and Stoffer}}]{Hoferichter:2023sli}
\bibinfo{author}{\bibfnamefont{M.}~\bibnamefont{Hoferichter}},
  \bibinfo{author}{\bibfnamefont{G.}~\bibnamefont{Colangelo}},
  \bibinfo{author}{\bibfnamefont{B.-L.} \bibnamefont{Hoid}},
  \bibinfo{author}{\bibfnamefont{B.}~\bibnamefont{Kubis}},
  \bibinfo{author}{\bibfnamefont{J.}~\bibnamefont{Ruiz~de Elvira}},
  \bibinfo{author}{\bibfnamefont{D.}~\bibnamefont{Schuh}},
  \bibinfo{author}{\bibfnamefont{D.}~\bibnamefont{Stamen}}, \bibnamefont{and}
  \bibinfo{author}{\bibfnamefont{P.}~\bibnamefont{Stoffer}},
  \bibinfo{journal}{Phys. Rev. Lett.} \textbf{\bibinfo{volume}{131}},
  \bibinfo{pages}{161905} (\bibinfo{year}{2023}{\natexlab{a}}),
  \eprint{2307.02532}.

\bibitem[{\citenamefont{Hoferichter
  et~al.}(2023{\natexlab{b}})\citenamefont{Hoferichter, Hoid, Kubis, and
  Schuh}}]{Hoferichter:2023bjm}
\bibinfo{author}{\bibfnamefont{M.}~\bibnamefont{Hoferichter}},
  \bibinfo{author}{\bibfnamefont{B.-L.} \bibnamefont{Hoid}},
  \bibinfo{author}{\bibfnamefont{B.}~\bibnamefont{Kubis}}, \bibnamefont{and}
  \bibinfo{author}{\bibfnamefont{D.}~\bibnamefont{Schuh}},
  \bibinfo{journal}{JHEP} \textbf{\bibinfo{volume}{08}}, \bibinfo{pages}{208}
  (\bibinfo{year}{2023}{\natexlab{b}}), \eprint{2307.02546}.

\bibitem[{\citenamefont{Stoffer et~al.}(2023)\citenamefont{Stoffer, Colangelo,
  and Hoferichter}}]{Stoffer:2023gba}
\bibinfo{author}{\bibfnamefont{P.}~\bibnamefont{Stoffer}},
  \bibinfo{author}{\bibfnamefont{G.}~\bibnamefont{Colangelo}},
  \bibnamefont{and}
  \bibinfo{author}{\bibfnamefont{M.}~\bibnamefont{Hoferichter}},
  \bibinfo{journal}{JINST} \textbf{\bibinfo{volume}{18}},
  \bibinfo{pages}{C10021} (\bibinfo{year}{2023}), \eprint{2308.04217}.

\bibitem[{\citenamefont{Davier et~al.}(2024)\citenamefont{Davier, Hoecker,
  Lutz, Malaescu, and Zhang}}]{Davier:2023fpl}
\bibinfo{author}{\bibfnamefont{M.}~\bibnamefont{Davier}},
  \bibinfo{author}{\bibfnamefont{A.}~\bibnamefont{Hoecker}},
  \bibinfo{author}{\bibfnamefont{A.-M.} \bibnamefont{Lutz}},
  \bibinfo{author}{\bibfnamefont{B.}~\bibnamefont{Malaescu}}, \bibnamefont{and}
  \bibinfo{author}{\bibfnamefont{Z.}~\bibnamefont{Zhang}},
  \bibinfo{journal}{Eur. Phys. J. C} \textbf{\bibinfo{volume}{84}},
  \bibinfo{pages}{721} (\bibinfo{year}{2024}), \eprint{2312.02053}.

\bibitem[{\citenamefont{Ignatov et~al.}(2024{\natexlab{a}})}]{CMD-3:2023alj}
\bibinfo{author}{\bibfnamefont{F.~V.} \bibnamefont{Ignatov}}
  \bibnamefont{et~al.} (\bibinfo{collaboration}{CMD-3}),
  \bibinfo{journal}{Phys. Rev. D} \textbf{\bibinfo{volume}{109}},
  \bibinfo{pages}{112002} (\bibinfo{year}{2024}{\natexlab{a}}),
  \eprint{2302.08834}.

\bibitem[{\citenamefont{Ignatov et~al.}(2024{\natexlab{b}})}]{CMD-3:2023rfe}
\bibinfo{author}{\bibfnamefont{F.~V.} \bibnamefont{Ignatov}}
  \bibnamefont{et~al.} (\bibinfo{collaboration}{CMD-3}),
  \bibinfo{journal}{Phys. Rev. Lett.} \textbf{\bibinfo{volume}{132}},
  \bibinfo{pages}{231903} (\bibinfo{year}{2024}{\natexlab{b}}),
  \eprint{2309.12910}.

\bibitem[{\citenamefont{Leplumey and Stoffer}(2025)}]{Leplumey:2025kvv}
\bibinfo{author}{\bibfnamefont{T.~P.} \bibnamefont{Leplumey}} \bibnamefont{and}
  \bibinfo{author}{\bibfnamefont{P.}~\bibnamefont{Stoffer}}
  (\bibinfo{year}{2025}), \eprint{2501.09643}.

\bibitem[{\citenamefont{Bors\'anyi et~al.}(2021)}]{Borsanyi:2020mff}
\bibinfo{author}{\bibfnamefont{S.}~\bibnamefont{Bors\'anyi}}
  \bibnamefont{et~al.} (\bibinfo{collaboration}{BMWc}),
  \bibinfo{journal}{Nature} \textbf{\bibinfo{volume}{593}}, \bibinfo{pages}{51}
  (\bibinfo{year}{2021}), \eprint{2002.12347}.

\bibitem[{\citenamefont{C\`e et~al.}(2022)}]{Ce:2022kxy}
\bibinfo{author}{\bibfnamefont{M.}~\bibnamefont{C\`e}} \bibnamefont{et~al.},
  \bibinfo{journal}{Phys. Rev. D} \textbf{\bibinfo{volume}{106}},
  \bibinfo{pages}{114502} (\bibinfo{year}{2022}), \eprint{2206.06582}.

\bibitem[{\citenamefont{Alexandrou et~al.}(2023)}]{ExtendedTwistedMass:2022jpw}
\bibinfo{author}{\bibfnamefont{C.}~\bibnamefont{Alexandrou}}
  \bibnamefont{et~al.} (\bibinfo{collaboration}{ETM}), \bibinfo{journal}{Phys.
  Rev. D} \textbf{\bibinfo{volume}{107}}, \bibinfo{pages}{074506}
  (\bibinfo{year}{2023}), \eprint{2206.15084}.

\bibitem[{\citenamefont{Bazavov et~al.}(2023)}]{FermilabLatticeHPQCD:2023jof}
\bibinfo{author}{\bibfnamefont{A.}~\bibnamefont{Bazavov}} \bibnamefont{et~al.}
  (\bibinfo{collaboration}{Fermilab Lattice, HPQCD, MILC}),
  \bibinfo{journal}{Phys. Rev. D} \textbf{\bibinfo{volume}{107}},
  \bibinfo{pages}{114514} (\bibinfo{year}{2023}), \eprint{2301.08274}.

\bibitem[{\citenamefont{Blum et~al.}(2023)}]{RBC:2023pvn}
\bibinfo{author}{\bibfnamefont{T.}~\bibnamefont{Blum}} \bibnamefont{et~al.}
  (\bibinfo{collaboration}{RBC, UKQCD}), \bibinfo{journal}{Phys. Rev. D}
  \textbf{\bibinfo{volume}{108}}, \bibinfo{pages}{054507}
  (\bibinfo{year}{2023}), \eprint{2301.08696}.

\bibitem[{\citenamefont{Boccaletti et~al.}(2024)}]{Boccaletti:2024guq}
\bibinfo{author}{\bibfnamefont{A.}~\bibnamefont{Boccaletti}}
  \bibnamefont{et~al.} (\bibinfo{collaboration}{BMWc}) (\bibinfo{year}{2024}),
  \eprint{2407.10913}.

\bibitem[{\citenamefont{Blum et~al.}(2024)}]{Blum:2024drk}
\bibinfo{author}{\bibfnamefont{T.}~\bibnamefont{Blum}} \bibnamefont{et~al.}
  (\bibinfo{collaboration}{RBC, UKQCD}) (\bibinfo{year}{2024}),
  \eprint{2410.20590}.

\bibitem[{\citenamefont{Djukanovic et~al.}(2024)\citenamefont{Djukanovic, von
  Hippel, Kuberski, Meyer, Miller, Ottnad, Parrino, Risch, and
  Wittig}}]{Djukanovic:2024cmq}
\bibinfo{author}{\bibfnamefont{D.}~\bibnamefont{Djukanovic}},
  \bibinfo{author}{\bibfnamefont{G.}~\bibnamefont{von Hippel}},
  \bibinfo{author}{\bibfnamefont{S.}~\bibnamefont{Kuberski}},
  \bibinfo{author}{\bibfnamefont{H.~B.} \bibnamefont{Meyer}},
  \bibinfo{author}{\bibfnamefont{N.}~\bibnamefont{Miller}},
  \bibinfo{author}{\bibfnamefont{K.}~\bibnamefont{Ottnad}},
  \bibinfo{author}{\bibfnamefont{J.}~\bibnamefont{Parrino}},
  \bibinfo{author}{\bibfnamefont{A.}~\bibnamefont{Risch}}, \bibnamefont{and}
  \bibinfo{author}{\bibfnamefont{H.}~\bibnamefont{Wittig}}
  (\bibinfo{year}{2024}), \eprint{2411.07969}.

\bibitem[{\citenamefont{Bazavov et~al.}(2024)}]{Bazavov:2024eou}
\bibinfo{author}{\bibfnamefont{A.}~\bibnamefont{Bazavov}} \bibnamefont{et~al.}
  (\bibinfo{collaboration}{Fermilab Lattice, HPQCD, MILC})
  (\bibinfo{year}{2024}), \eprint{2412.18491}.

\bibitem[{\citenamefont{Colangelo
  et~al.}(2022{\natexlab{c}})}]{Colangelo:2022jxc}
\bibinfo{author}{\bibfnamefont{G.}~\bibnamefont{Colangelo}}
  \bibnamefont{et~al.} (\bibinfo{year}{2022}{\natexlab{c}}),
  \eprint{2203.15810}.

\bibitem[{\citenamefont{Campanario et~al.}(2019)\citenamefont{Campanario,
  Czy\.z, Gluza, Jeli\'nski, Rodrigo, Tracz, and
  Zhuridov}}]{Campanario:2019mjh}
\bibinfo{author}{\bibfnamefont{F.}~\bibnamefont{Campanario}},
  \bibinfo{author}{\bibfnamefont{H.}~\bibnamefont{Czy\.z}},
  \bibinfo{author}{\bibfnamefont{J.}~\bibnamefont{Gluza}},
  \bibinfo{author}{\bibfnamefont{T.}~\bibnamefont{Jeli\'nski}},
  \bibinfo{author}{\bibfnamefont{G.}~\bibnamefont{Rodrigo}},
  \bibinfo{author}{\bibfnamefont{S.}~\bibnamefont{Tracz}}, \bibnamefont{and}
  \bibinfo{author}{\bibfnamefont{D.}~\bibnamefont{Zhuridov}},
  \bibinfo{journal}{Phys. Rev. D} \textbf{\bibinfo{volume}{100}},
  \bibinfo{pages}{076004} (\bibinfo{year}{2019}), \eprint{1903.10197}.

\bibitem[{\citenamefont{Ignatov and Lee}(2022)}]{Ignatov:2022iou}
\bibinfo{author}{\bibfnamefont{F.}~\bibnamefont{Ignatov}} \bibnamefont{and}
  \bibinfo{author}{\bibfnamefont{R.~N.} \bibnamefont{Lee}},
  \bibinfo{journal}{Phys. Lett. B} \textbf{\bibinfo{volume}{833}},
  \bibinfo{pages}{137283} (\bibinfo{year}{2022}), \eprint{2204.12235}.

\bibitem[{\citenamefont{Colangelo
  et~al.}(2022{\natexlab{d}})\citenamefont{Colangelo, Hoferichter, Monnard, and
  Ruiz~de Elvira}}]{Colangelo:2022lzg}
\bibinfo{author}{\bibfnamefont{G.}~\bibnamefont{Colangelo}},
  \bibinfo{author}{\bibfnamefont{M.}~\bibnamefont{Hoferichter}},
  \bibinfo{author}{\bibfnamefont{J.}~\bibnamefont{Monnard}}, \bibnamefont{and}
  \bibinfo{author}{\bibfnamefont{J.}~\bibnamefont{Ruiz~de Elvira}},
  \bibinfo{journal}{JHEP} \textbf{\bibinfo{volume}{08}}, \bibinfo{pages}{295}
  (\bibinfo{year}{2022}{\natexlab{d}}), \bibinfo{note}{[Erratum: JHEP {\bf 09},
  177 (2024)]}, \eprint{2207.03495}.

\bibitem[{\citenamefont{Monnard}(2021)}]{Monnard:2021pvm}
\bibinfo{author}{\bibfnamefont{J.}~\bibnamefont{Monnard}}, Ph.D. thesis,
  \bibinfo{school}{Universit{\"a}t Bern} (\bibinfo{year}{2021}),
  \urlprefix\url{https://boristheses.unibe.ch/2825/}.

\bibitem[{\citenamefont{Abbiendi et~al.}(2022)}]{Abbiendi:2022liz}
\bibinfo{author}{\bibfnamefont{G.}~\bibnamefont{Abbiendi}} \bibnamefont{et~al.}
  (\bibinfo{year}{2022}), \eprint{2201.12102}.

\bibitem[{\citenamefont{Lees et~al.}(2023{\natexlab{a}})}]{BaBar:2023xiy}
\bibinfo{author}{\bibfnamefont{J.~P.} \bibnamefont{Lees}} \bibnamefont{et~al.}
  (\bibinfo{collaboration}{BaBar}), \bibinfo{journal}{Phys. Rev. D}
  \textbf{\bibinfo{volume}{108}}, \bibinfo{pages}{L111103}
  (\bibinfo{year}{2023}{\natexlab{a}}), \eprint{2308.05233}.

\bibitem[{\citenamefont{Aliberti et~al.}(2024)}]{Aliberti:2024fpq}
\bibinfo{author}{\bibfnamefont{R.}~\bibnamefont{Aliberti}} \bibnamefont{et~al.}
  (\bibinfo{year}{2024}), \eprint{2410.22882}.

\bibitem[{\citenamefont{Colangelo
  et~al.}(2014{\natexlab{b}})\citenamefont{Colangelo, Hoferichter, Procura, and
  Stoffer}}]{Colangelo:2014dfa}
\bibinfo{author}{\bibfnamefont{G.}~\bibnamefont{Colangelo}},
  \bibinfo{author}{\bibfnamefont{M.}~\bibnamefont{Hoferichter}},
  \bibinfo{author}{\bibfnamefont{M.}~\bibnamefont{Procura}}, \bibnamefont{and}
  \bibinfo{author}{\bibfnamefont{P.}~\bibnamefont{Stoffer}},
  \bibinfo{journal}{JHEP} \textbf{\bibinfo{volume}{09}}, \bibinfo{pages}{091}
  (\bibinfo{year}{2014}{\natexlab{b}}), \eprint{1402.7081}.

\bibitem[{\citenamefont{Colangelo
  et~al.}(2014{\natexlab{c}})\citenamefont{Colangelo, Hoferichter, Kubis,
  Procura, and Stoffer}}]{Colangelo:2014pva}
\bibinfo{author}{\bibfnamefont{G.}~\bibnamefont{Colangelo}},
  \bibinfo{author}{\bibfnamefont{M.}~\bibnamefont{Hoferichter}},
  \bibinfo{author}{\bibfnamefont{B.}~\bibnamefont{Kubis}},
  \bibinfo{author}{\bibfnamefont{M.}~\bibnamefont{Procura}}, \bibnamefont{and}
  \bibinfo{author}{\bibfnamefont{P.}~\bibnamefont{Stoffer}},
  \bibinfo{journal}{Phys. Lett. B} \textbf{\bibinfo{volume}{738}},
  \bibinfo{pages}{6} (\bibinfo{year}{2014}{\natexlab{c}}), \eprint{1408.2517}.

\bibitem[{\citenamefont{Colangelo et~al.}(2015)\citenamefont{Colangelo,
  Hoferichter, Procura, and Stoffer}}]{Colangelo:2015ama}
\bibinfo{author}{\bibfnamefont{G.}~\bibnamefont{Colangelo}},
  \bibinfo{author}{\bibfnamefont{M.}~\bibnamefont{Hoferichter}},
  \bibinfo{author}{\bibfnamefont{M.}~\bibnamefont{Procura}}, \bibnamefont{and}
  \bibinfo{author}{\bibfnamefont{P.}~\bibnamefont{Stoffer}},
  \bibinfo{journal}{JHEP} \textbf{\bibinfo{volume}{09}}, \bibinfo{pages}{074}
  (\bibinfo{year}{2015}), \eprint{1506.01386}.

\bibitem[{\citenamefont{Hoferichter
  et~al.}(2014{\natexlab{a}})\citenamefont{Hoferichter, Colangelo, Procura, and
  Stoffer}}]{Hoferichter:2013ama}
\bibinfo{author}{\bibfnamefont{M.}~\bibnamefont{Hoferichter}},
  \bibinfo{author}{\bibfnamefont{G.}~\bibnamefont{Colangelo}},
  \bibinfo{author}{\bibfnamefont{M.}~\bibnamefont{Procura}}, \bibnamefont{and}
  \bibinfo{author}{\bibfnamefont{P.}~\bibnamefont{Stoffer}},
  \bibinfo{journal}{Int. J. Mod. Phys. Conf. Ser.}
  \textbf{\bibinfo{volume}{35}}, \bibinfo{pages}{1460400}
  (\bibinfo{year}{2014}{\natexlab{a}}), \eprint{1309.6877}.

\bibitem[{\citenamefont{Zanke et~al.}(2021)\citenamefont{Zanke, Hoferichter,
  and Kubis}}]{Zanke:2021wiq}
\bibinfo{author}{\bibfnamefont{M.}~\bibnamefont{Zanke}},
  \bibinfo{author}{\bibfnamefont{M.}~\bibnamefont{Hoferichter}},
  \bibnamefont{and} \bibinfo{author}{\bibfnamefont{B.}~\bibnamefont{Kubis}},
  \bibinfo{journal}{JHEP} \textbf{\bibinfo{volume}{07}}, \bibinfo{pages}{106}
  (\bibinfo{year}{2021}), \eprint{2103.09829}.

\bibitem[{\citenamefont{Hoferichter
  et~al.}(2023{\natexlab{c}})\citenamefont{Hoferichter, Kubis, and
  Zanke}}]{Hoferichter:2023tgp}
\bibinfo{author}{\bibfnamefont{M.}~\bibnamefont{Hoferichter}},
  \bibinfo{author}{\bibfnamefont{B.}~\bibnamefont{Kubis}}, \bibnamefont{and}
  \bibinfo{author}{\bibfnamefont{M.}~\bibnamefont{Zanke}},
  \bibinfo{journal}{JHEP} \textbf{\bibinfo{volume}{08}}, \bibinfo{pages}{209}
  (\bibinfo{year}{2023}{\natexlab{c}}), \eprint{2307.14413}.

\bibitem[{\citenamefont{Hoferichter and Stoffer}(2020)}]{Hoferichter:2020lap}
\bibinfo{author}{\bibfnamefont{M.}~\bibnamefont{Hoferichter}} \bibnamefont{and}
  \bibinfo{author}{\bibfnamefont{P.}~\bibnamefont{Stoffer}},
  \bibinfo{journal}{JHEP} \textbf{\bibinfo{volume}{05}}, \bibinfo{pages}{159}
  (\bibinfo{year}{2020}), \eprint{2004.06127}.

\bibitem[{\citenamefont{Hoferichter
  et~al.}(2024{\natexlab{a}})\citenamefont{Hoferichter, Stoffer, and
  Zillinger}}]{Hoferichter:2024fsj}
\bibinfo{author}{\bibfnamefont{M.}~\bibnamefont{Hoferichter}},
  \bibinfo{author}{\bibfnamefont{P.}~\bibnamefont{Stoffer}}, \bibnamefont{and}
  \bibinfo{author}{\bibfnamefont{M.}~\bibnamefont{Zillinger}},
  \bibinfo{journal}{JHEP} \textbf{\bibinfo{volume}{04}}, \bibinfo{pages}{092}
  (\bibinfo{year}{2024}{\natexlab{a}}), \eprint{2402.14060}.

\bibitem[{\citenamefont{Bijnens et~al.}(2020)\citenamefont{Bijnens,
  Hermansson-Truedsson, Laub, and Rodr\'iguez-S\'anchez}}]{Bijnens:2020xnl}
\bibinfo{author}{\bibfnamefont{J.}~\bibnamefont{Bijnens}},
  \bibinfo{author}{\bibfnamefont{N.}~\bibnamefont{Hermansson-Truedsson}},
  \bibinfo{author}{\bibfnamefont{L.}~\bibnamefont{Laub}}, \bibnamefont{and}
  \bibinfo{author}{\bibfnamefont{A.}~\bibnamefont{Rodr\'iguez-S\'anchez}},
  \bibinfo{journal}{JHEP} \textbf{\bibinfo{volume}{10}}, \bibinfo{pages}{203}
  (\bibinfo{year}{2020}), \eprint{2008.13487}.

\bibitem[{\citenamefont{Bijnens et~al.}(2021)\citenamefont{Bijnens,
  Hermansson-Truedsson, Laub, and Rodr\'iguez-S\'anchez}}]{Bijnens:2021jqo}
\bibinfo{author}{\bibfnamefont{J.}~\bibnamefont{Bijnens}},
  \bibinfo{author}{\bibfnamefont{N.}~\bibnamefont{Hermansson-Truedsson}},
  \bibinfo{author}{\bibfnamefont{L.}~\bibnamefont{Laub}}, \bibnamefont{and}
  \bibinfo{author}{\bibfnamefont{A.}~\bibnamefont{Rodr\'iguez-S\'anchez}},
  \bibinfo{journal}{JHEP} \textbf{\bibinfo{volume}{04}}, \bibinfo{pages}{240}
  (\bibinfo{year}{2021}), \eprint{2101.09169}.

\bibitem[{\citenamefont{Bijnens et~al.}(2023)\citenamefont{Bijnens,
  Hermansson-Truedsson, and Rodr\'\i{}guez-S\'anchez}}]{Bijnens:2022itw}
\bibinfo{author}{\bibfnamefont{J.}~\bibnamefont{Bijnens}},
  \bibinfo{author}{\bibfnamefont{N.}~\bibnamefont{Hermansson-Truedsson}},
  \bibnamefont{and}
  \bibinfo{author}{\bibfnamefont{A.}~\bibnamefont{Rodr\'\i{}guez-S\'anchez}},
  \bibinfo{journal}{JHEP} \textbf{\bibinfo{volume}{02}}, \bibinfo{pages}{167}
  (\bibinfo{year}{2023}), \eprint{2211.17183}.

\bibitem[{\citenamefont{Bijnens et~al.}(2024)\citenamefont{Bijnens,
  Hermansson-Truedsson, and Rodr\'\i{}guez-S\'anchez}}]{Bijnens:2024jgh}
\bibinfo{author}{\bibfnamefont{J.}~\bibnamefont{Bijnens}},
  \bibinfo{author}{\bibfnamefont{N.}~\bibnamefont{Hermansson-Truedsson}},
  \bibnamefont{and}
  \bibinfo{author}{\bibfnamefont{A.}~\bibnamefont{Rodr\'\i{}guez-S\'anchez}}
  (\bibinfo{year}{2024}), \eprint{2411.09578}.

\bibitem[{\citenamefont{L\"udtke et~al.}(2024)\citenamefont{L\"udtke, Procura,
  and Stoffer}}]{Ludtke:2024ase}
\bibinfo{author}{\bibfnamefont{J.}~\bibnamefont{L\"udtke}},
  \bibinfo{author}{\bibfnamefont{M.}~\bibnamefont{Procura}}, \bibnamefont{and}
  \bibinfo{author}{\bibfnamefont{P.}~\bibnamefont{Stoffer}}
  (\bibinfo{year}{2024}), \eprint{2410.11946}.

\bibitem[{\citenamefont{Hoferichter
  et~al.}(2024{\natexlab{b}})\citenamefont{Hoferichter, Stoffer, and
  Zillinger}}]{Hoferichter:2024bae}
\bibinfo{author}{\bibfnamefont{M.}~\bibnamefont{Hoferichter}},
  \bibinfo{author}{\bibfnamefont{P.}~\bibnamefont{Stoffer}}, \bibnamefont{and}
  \bibinfo{author}{\bibfnamefont{M.}~\bibnamefont{Zillinger}}
  (\bibinfo{year}{2024}{\natexlab{b}}), \eprint{2412.00178}.

\bibitem[{\citenamefont{Bardeen and Tung}(1968)}]{Bardeen:1968ebo}
\bibinfo{author}{\bibfnamefont{W.~A.} \bibnamefont{Bardeen}} \bibnamefont{and}
  \bibinfo{author}{\bibfnamefont{W.~K.} \bibnamefont{Tung}},
  \bibinfo{journal}{Phys. Rev.} \textbf{\bibinfo{volume}{173}},
  \bibinfo{pages}{1423} (\bibinfo{year}{1968}), \bibinfo{note}{[Erratum: Phys.
  Rev. D {\bf 4}, 3229 (1971)]}.

\bibitem[{\citenamefont{Tarrach}(1975)}]{Tarrach:1975tu}
\bibinfo{author}{\bibfnamefont{R.}~\bibnamefont{Tarrach}},
  \bibinfo{journal}{Nuovo Cim. A} \textbf{\bibinfo{volume}{28}},
  \bibinfo{pages}{409} (\bibinfo{year}{1975}).

\bibitem[{\citenamefont{Eichmann et~al.}(2015)\citenamefont{Eichmann, Fischer,
  and Heupel}}]{Eichmann:2015nra}
\bibinfo{author}{\bibfnamefont{G.}~\bibnamefont{Eichmann}},
  \bibinfo{author}{\bibfnamefont{C.~S.} \bibnamefont{Fischer}},
  \bibnamefont{and} \bibinfo{author}{\bibfnamefont{W.}~\bibnamefont{Heupel}},
  \bibinfo{journal}{Phys. Rev. D} \textbf{\bibinfo{volume}{92}},
  \bibinfo{pages}{056006} (\bibinfo{year}{2015}), \eprint{1505.06336}.

\bibitem[{\citenamefont{L\"udtke et~al.}(2023)\citenamefont{L\"udtke, Procura,
  and Stoffer}}]{Ludtke:2023hvz}
\bibinfo{author}{\bibfnamefont{J.}~\bibnamefont{L\"udtke}},
  \bibinfo{author}{\bibfnamefont{M.}~\bibnamefont{Procura}}, \bibnamefont{and}
  \bibinfo{author}{\bibfnamefont{P.}~\bibnamefont{Stoffer}},
  \bibinfo{journal}{JHEP} \textbf{\bibinfo{volume}{04}}, \bibinfo{pages}{125}
  (\bibinfo{year}{2023}), \eprint{2302.12264}.

\bibitem[{\citenamefont{Schneider et~al.}(2012)\citenamefont{Schneider, Kubis,
  and Niecknig}}]{Schneider:2012ez}
\bibinfo{author}{\bibfnamefont{S.~P.} \bibnamefont{Schneider}},
  \bibinfo{author}{\bibfnamefont{B.}~\bibnamefont{Kubis}}, \bibnamefont{and}
  \bibinfo{author}{\bibfnamefont{F.}~\bibnamefont{Niecknig}},
  \bibinfo{journal}{Phys. Rev. D} \textbf{\bibinfo{volume}{86}},
  \bibinfo{pages}{054013} (\bibinfo{year}{2012}), \eprint{1206.3098}.

\bibitem[{\citenamefont{Hoferichter et~al.}(2012)\citenamefont{Hoferichter,
  Kubis, and Sakkas}}]{Hoferichter:2012pm}
\bibinfo{author}{\bibfnamefont{M.}~\bibnamefont{Hoferichter}},
  \bibinfo{author}{\bibfnamefont{B.}~\bibnamefont{Kubis}}, \bibnamefont{and}
  \bibinfo{author}{\bibfnamefont{D.}~\bibnamefont{Sakkas}},
  \bibinfo{journal}{Phys. Rev. D} \textbf{\bibinfo{volume}{86}},
  \bibinfo{pages}{116009} (\bibinfo{year}{2012}), \eprint{1210.6793}.

\bibitem[{\citenamefont{Hoferichter
  et~al.}(2014{\natexlab{b}})\citenamefont{Hoferichter, Kubis, Leupold,
  Niecknig, and Schneider}}]{Hoferichter:2014vra}
\bibinfo{author}{\bibfnamefont{M.}~\bibnamefont{Hoferichter}},
  \bibinfo{author}{\bibfnamefont{B.}~\bibnamefont{Kubis}},
  \bibinfo{author}{\bibfnamefont{S.}~\bibnamefont{Leupold}},
  \bibinfo{author}{\bibfnamefont{F.}~\bibnamefont{Niecknig}}, \bibnamefont{and}
  \bibinfo{author}{\bibfnamefont{S.~P.} \bibnamefont{Schneider}},
  \bibinfo{journal}{Eur. Phys. J. C} \textbf{\bibinfo{volume}{74}},
  \bibinfo{pages}{3180} (\bibinfo{year}{2014}{\natexlab{b}}),
  \eprint{1410.4691}.

\bibitem[{\citenamefont{Hoferichter et~al.}(2022)\citenamefont{Hoferichter,
  Hoid, Kubis, and L\"udtke}}]{Hoferichter:2021lct}
\bibinfo{author}{\bibfnamefont{M.}~\bibnamefont{Hoferichter}},
  \bibinfo{author}{\bibfnamefont{B.-L.} \bibnamefont{Hoid}},
  \bibinfo{author}{\bibfnamefont{B.}~\bibnamefont{Kubis}}, \bibnamefont{and}
  \bibinfo{author}{\bibfnamefont{J.}~\bibnamefont{L\"udtke}},
  \bibinfo{journal}{Phys. Rev. Lett.} \textbf{\bibinfo{volume}{128}},
  \bibinfo{pages}{172004} (\bibinfo{year}{2022}), \eprint{2105.04563}.

\bibitem[{\citenamefont{Stollenwerk et~al.}(2012)\citenamefont{Stollenwerk,
  Hanhart, Kup\'s\'c, Mei{\ss}ner, and Wirzba}}]{Stollenwerk:2011zz}
\bibinfo{author}{\bibfnamefont{F.}~\bibnamefont{Stollenwerk}},
  \bibinfo{author}{\bibfnamefont{C.}~\bibnamefont{Hanhart}},
  \bibinfo{author}{\bibfnamefont{A.}~\bibnamefont{Kup\'s\'c}},
  \bibinfo{author}{\bibfnamefont{U.-G.} \bibnamefont{Mei{\ss}ner}},
  \bibnamefont{and} \bibinfo{author}{\bibfnamefont{A.}~\bibnamefont{Wirzba}},
  \bibinfo{journal}{Phys. Lett. B} \textbf{\bibinfo{volume}{707}},
  \bibinfo{pages}{184} (\bibinfo{year}{2012}), \eprint{1108.2419}.

\bibitem[{\citenamefont{Hanhart et~al.}(2013)\citenamefont{Hanhart, Kup\'s\'c,
  Mei\ss{}ner, Stollenwerk, and Wirzba}}]{Hanhart:2013vba}
\bibinfo{author}{\bibfnamefont{C.}~\bibnamefont{Hanhart}},
  \bibinfo{author}{\bibfnamefont{A.}~\bibnamefont{Kup\'s\'c}},
  \bibinfo{author}{\bibfnamefont{U.-G.} \bibnamefont{Mei\ss{}ner}},
  \bibinfo{author}{\bibfnamefont{F.}~\bibnamefont{Stollenwerk}},
  \bibnamefont{and} \bibinfo{author}{\bibfnamefont{A.}~\bibnamefont{Wirzba}},
  \bibinfo{journal}{Eur. Phys. J. C} \textbf{\bibinfo{volume}{73}},
  \bibinfo{pages}{2668} (\bibinfo{year}{2013}), \bibinfo{note}{[Erratum: Eur.
  Phys. J. C \textbf{75}, 242 (2015)]}, \eprint{1307.5654}.

\bibitem[{\citenamefont{Kubis and Plenter}(2015)}]{Kubis:2015sga}
\bibinfo{author}{\bibfnamefont{B.}~\bibnamefont{Kubis}} \bibnamefont{and}
  \bibinfo{author}{\bibfnamefont{J.}~\bibnamefont{Plenter}},
  \bibinfo{journal}{Eur. Phys. J. C} \textbf{\bibinfo{volume}{75}},
  \bibinfo{pages}{283} (\bibinfo{year}{2015}), \eprint{1504.02588}.

\bibitem[{\citenamefont{Holz et~al.}(2021)\citenamefont{Holz, Plenter, Xiao,
  Dato, Hanhart, Kubis, Mei\ss{}ner, and Wirzba}}]{Holz:2015tcg}
\bibinfo{author}{\bibfnamefont{S.}~\bibnamefont{Holz}},
  \bibinfo{author}{\bibfnamefont{J.}~\bibnamefont{Plenter}},
  \bibinfo{author}{\bibfnamefont{C.-W.} \bibnamefont{Xiao}},
  \bibinfo{author}{\bibfnamefont{T.}~\bibnamefont{Dato}},
  \bibinfo{author}{\bibfnamefont{C.}~\bibnamefont{Hanhart}},
  \bibinfo{author}{\bibfnamefont{B.}~\bibnamefont{Kubis}},
  \bibinfo{author}{\bibfnamefont{U.-G.} \bibnamefont{Mei\ss{}ner}},
  \bibnamefont{and} \bibinfo{author}{\bibfnamefont{A.}~\bibnamefont{Wirzba}},
  \bibinfo{journal}{Eur. Phys. J. C} \textbf{\bibinfo{volume}{81}},
  \bibinfo{pages}{1002} (\bibinfo{year}{2021}), \eprint{1509.02194}.

\bibitem[{\citenamefont{Holz et~al.}(2022)\citenamefont{Holz, Hanhart,
  Hoferichter, and Kubis}}]{Holz:2022hwz}
\bibinfo{author}{\bibfnamefont{S.}~\bibnamefont{Holz}},
  \bibinfo{author}{\bibfnamefont{C.}~\bibnamefont{Hanhart}},
  \bibinfo{author}{\bibfnamefont{M.}~\bibnamefont{Hoferichter}},
  \bibnamefont{and} \bibinfo{author}{\bibfnamefont{B.}~\bibnamefont{Kubis}},
  \bibinfo{journal}{Eur. Phys. J. C} \textbf{\bibinfo{volume}{82}},
  \bibinfo{pages}{434} (\bibinfo{year}{2022}), \bibinfo{note}{[Addendum: Eur.
  Phys. J. C {\bf 82}, 1159 (2022)]}, \eprint{2202.05846}.

\bibitem[{\citenamefont{Holz}(2022)}]{Holz:2022smu}
\bibinfo{author}{\bibfnamefont{S.}~\bibnamefont{Holz}}, Ph.D. thesis,
  \bibinfo{school}{University of Bonn} (\bibinfo{year}{2022}),
  \urlprefix\url{https://nbn-resolving.org/urn:nbn:de:hbz:5-67976}.

\bibitem[{\citenamefont{Holz et~al.}(2024{\natexlab{a}})\citenamefont{Holz,
  Hoferichter, Hoid, and Kubis}}]{Holz:2024lom}
\bibinfo{author}{\bibfnamefont{S.}~\bibnamefont{Holz}},
  \bibinfo{author}{\bibfnamefont{M.}~\bibnamefont{Hoferichter}},
  \bibinfo{author}{\bibfnamefont{B.-L.} \bibnamefont{Hoid}}, \bibnamefont{and}
  \bibinfo{author}{\bibfnamefont{B.}~\bibnamefont{Kubis}}
  (\bibinfo{year}{2024}{\natexlab{a}}), \eprint{2411.08098}.

\bibitem[{\citenamefont{Holz et~al.}(2024{\natexlab{b}})\citenamefont{Holz,
  Hoferichter, Hoid, and Kubis}}]{Holz:2024diw}
\bibinfo{author}{\bibfnamefont{S.}~\bibnamefont{Holz}},
  \bibinfo{author}{\bibfnamefont{M.}~\bibnamefont{Hoferichter}},
  \bibinfo{author}{\bibfnamefont{B.-L.} \bibnamefont{Hoid}}, \bibnamefont{and}
  \bibinfo{author}{\bibfnamefont{B.}~\bibnamefont{Kubis}}
  (\bibinfo{year}{2024}{\natexlab{b}}), \eprint{2412.16281}.

\bibitem[{\citenamefont{Garc\'ia-Mart\'in and
  Moussallam}(2010)}]{Garcia-Martin:2010kyn}
\bibinfo{author}{\bibfnamefont{R.}~\bibnamefont{Garc\'ia-Mart\'in}}
  \bibnamefont{and}
  \bibinfo{author}{\bibfnamefont{B.}~\bibnamefont{Moussallam}},
  \bibinfo{journal}{Eur. Phys. J. C} \textbf{\bibinfo{volume}{70}},
  \bibinfo{pages}{155} (\bibinfo{year}{2010}), \eprint{1006.5373}.

\bibitem[{\citenamefont{Hoferichter et~al.}(2011)\citenamefont{Hoferichter,
  Phillips, and Schat}}]{Hoferichter:2011wk}
\bibinfo{author}{\bibfnamefont{M.}~\bibnamefont{Hoferichter}},
  \bibinfo{author}{\bibfnamefont{D.~R.} \bibnamefont{Phillips}},
  \bibnamefont{and} \bibinfo{author}{\bibfnamefont{C.}~\bibnamefont{Schat}},
  \bibinfo{journal}{Eur. Phys. J. C} \textbf{\bibinfo{volume}{71}},
  \bibinfo{pages}{1743} (\bibinfo{year}{2011}), \eprint{1106.4147}.

\bibitem[{\citenamefont{Moussallam}(2013)}]{Moussallam:2013una}
\bibinfo{author}{\bibfnamefont{B.}~\bibnamefont{Moussallam}},
  \bibinfo{journal}{Eur. Phys. J. C} \textbf{\bibinfo{volume}{73}},
  \bibinfo{pages}{2539} (\bibinfo{year}{2013}), \eprint{1305.3143}.

\bibitem[{\citenamefont{Danilkin and Vanderhaeghen}(2019)}]{Danilkin:2018qfn}
\bibinfo{author}{\bibfnamefont{I.}~\bibnamefont{Danilkin}} \bibnamefont{and}
  \bibinfo{author}{\bibfnamefont{M.}~\bibnamefont{Vanderhaeghen}},
  \bibinfo{journal}{Phys. Lett. B} \textbf{\bibinfo{volume}{789}},
  \bibinfo{pages}{366} (\bibinfo{year}{2019}), \eprint{1810.03669}.

\bibitem[{\citenamefont{Hoferichter and Stoffer}(2019)}]{Hoferichter:2019nlq}
\bibinfo{author}{\bibfnamefont{M.}~\bibnamefont{Hoferichter}} \bibnamefont{and}
  \bibinfo{author}{\bibfnamefont{P.}~\bibnamefont{Stoffer}},
  \bibinfo{journal}{JHEP} \textbf{\bibinfo{volume}{07}}, \bibinfo{pages}{073}
  (\bibinfo{year}{2019}), \eprint{1905.13198}.

\bibitem[{\citenamefont{Danilkin et~al.}(2020)\citenamefont{Danilkin, Deineka,
  and Vanderhaeghen}}]{Danilkin:2019opj}
\bibinfo{author}{\bibfnamefont{I.}~\bibnamefont{Danilkin}},
  \bibinfo{author}{\bibfnamefont{O.}~\bibnamefont{Deineka}}, \bibnamefont{and}
  \bibinfo{author}{\bibfnamefont{M.}~\bibnamefont{Vanderhaeghen}},
  \bibinfo{journal}{Phys. Rev. D} \textbf{\bibinfo{volume}{101}},
  \bibinfo{pages}{054008} (\bibinfo{year}{2020}), \eprint{1909.04158}.

\bibitem[{\citenamefont{Lu and Moussallam}(2020)}]{Lu:2020qeo}
\bibinfo{author}{\bibfnamefont{J.}~\bibnamefont{Lu}} \bibnamefont{and}
  \bibinfo{author}{\bibfnamefont{B.}~\bibnamefont{Moussallam}},
  \bibinfo{journal}{Eur. Phys. J. C} \textbf{\bibinfo{volume}{80}},
  \bibinfo{pages}{436} (\bibinfo{year}{2020}), \eprint{2002.04441}.

\bibitem[{\citenamefont{Sch\"afer et~al.}(2023)\citenamefont{Sch\"afer, Zanke,
  Korte, and Kubis}}]{Schafer:2023qtl}
\bibinfo{author}{\bibfnamefont{H.}~\bibnamefont{Sch\"afer}},
  \bibinfo{author}{\bibfnamefont{M.}~\bibnamefont{Zanke}},
  \bibinfo{author}{\bibfnamefont{Y.}~\bibnamefont{Korte}}, \bibnamefont{and}
  \bibinfo{author}{\bibfnamefont{B.}~\bibnamefont{Kubis}},
  \bibinfo{journal}{Phys. Rev. D} \textbf{\bibinfo{volume}{108}},
  \bibinfo{pages}{074025} (\bibinfo{year}{2023}), \eprint{2307.10357}.

\bibitem[{\citenamefont{Deineka et~al.}(2024)\citenamefont{Deineka, Danilkin,
  and Vanderhaeghen}}]{Deineka:2024mzt}
\bibinfo{author}{\bibfnamefont{O.}~\bibnamefont{Deineka}},
  \bibinfo{author}{\bibfnamefont{I.}~\bibnamefont{Danilkin}}, \bibnamefont{and}
  \bibinfo{author}{\bibfnamefont{M.}~\bibnamefont{Vanderhaeghen}}
  (\bibinfo{year}{2024}), \eprint{2410.12894}.

\bibitem[{\citenamefont{Danilkin et~al.}(2021)\citenamefont{Danilkin,
  Hoferichter, and Stoffer}}]{Danilkin:2021icn}
\bibinfo{author}{\bibfnamefont{I.}~\bibnamefont{Danilkin}},
  \bibinfo{author}{\bibfnamefont{M.}~\bibnamefont{Hoferichter}},
  \bibnamefont{and} \bibinfo{author}{\bibfnamefont{P.}~\bibnamefont{Stoffer}},
  \bibinfo{journal}{Phys. Lett. B} \textbf{\bibinfo{volume}{820}},
  \bibinfo{pages}{136502} (\bibinfo{year}{2021}), \eprint{2105.01666}.

\bibitem[{\citenamefont{Achard et~al.}(2002)}]{Achard:2001uu}
\bibinfo{author}{\bibfnamefont{P.}~\bibnamefont{Achard}} \bibnamefont{et~al.}
  (\bibinfo{collaboration}{L3}), \bibinfo{journal}{Phys. Lett. B}
  \textbf{\bibinfo{volume}{526}}, \bibinfo{pages}{269} (\bibinfo{year}{2002}),
  \eprint{hep-ex/0110073}.

\bibitem[{\citenamefont{Achard et~al.}(2007)}]{Achard:2007hm}
\bibinfo{author}{\bibfnamefont{P.}~\bibnamefont{Achard}} \bibnamefont{et~al.}
  (\bibinfo{collaboration}{L3}), \bibinfo{journal}{JHEP}
  \textbf{\bibinfo{volume}{03}}, \bibinfo{pages}{018} (\bibinfo{year}{2007}).

\bibitem[{\citenamefont{Aubert et~al.}(2007)}]{BaBar:2007qju}
\bibinfo{author}{\bibfnamefont{B.}~\bibnamefont{Aubert}} \bibnamefont{et~al.}
  (\bibinfo{collaboration}{BaBar}), \bibinfo{journal}{Phys. Rev. D}
  \textbf{\bibinfo{volume}{76}}, \bibinfo{pages}{092005}
  (\bibinfo{year}{2007}), \bibinfo{note}{[Erratum: Phys. Rev. D \textbf{77},
  119902 (2008)]}, \eprint{0708.2461}.

\bibitem[{\citenamefont{Lees et~al.}(2023{\natexlab{b}})}]{BaBar:2022ahi}
\bibinfo{author}{\bibfnamefont{J.~P.} \bibnamefont{Lees}} \bibnamefont{et~al.}
  (\bibinfo{collaboration}{BaBar}), \bibinfo{journal}{Phys. Rev. D}
  \textbf{\bibinfo{volume}{107}}, \bibinfo{pages}{072001}
  (\bibinfo{year}{2023}{\natexlab{b}}), \eprint{2207.10340}.

\bibitem[{\citenamefont{Navas et~al.}(2024)}]{ParticleDataGroup:2024cfk}
\bibinfo{author}{\bibfnamefont{S.}~\bibnamefont{Navas}} \bibnamefont{et~al.}
  (\bibinfo{collaboration}{Particle Data Group}), \bibinfo{journal}{Phys. Rev.
  D} \textbf{\bibinfo{volume}{110}}, \bibinfo{pages}{030001}
  (\bibinfo{year}{2024}).

\bibitem[{\citenamefont{Leutgeb et~al.}(2023)\citenamefont{Leutgeb, Mager, and
  Rebhan}}]{Leutgeb:2022lqw}
\bibinfo{author}{\bibfnamefont{J.}~\bibnamefont{Leutgeb}},
  \bibinfo{author}{\bibfnamefont{J.}~\bibnamefont{Mager}}, \bibnamefont{and}
  \bibinfo{author}{\bibfnamefont{A.}~\bibnamefont{Rebhan}},
  \bibinfo{journal}{Phys. Rev. D} \textbf{\bibinfo{volume}{107}},
  \bibinfo{pages}{054021} (\bibinfo{year}{2023}), \eprint{2211.16562}.

\bibitem[{\citenamefont{Leutgeb et~al.}(2024)\citenamefont{Leutgeb, Mager, and
  Rebhan}}]{Leutgeb:2024rfs}
\bibinfo{author}{\bibfnamefont{J.}~\bibnamefont{Leutgeb}},
  \bibinfo{author}{\bibfnamefont{J.}~\bibnamefont{Mager}}, \bibnamefont{and}
  \bibinfo{author}{\bibfnamefont{A.}~\bibnamefont{Rebhan}}
  (\bibinfo{year}{2024}), \eprint{2411.10432}.

\bibitem[{\citenamefont{Schuler et~al.}(1998)\citenamefont{Schuler, Berends,
  and van Gulik}}]{Schuler:1997yw}
\bibinfo{author}{\bibfnamefont{G.~A.} \bibnamefont{Schuler}},
  \bibinfo{author}{\bibfnamefont{F.~A.} \bibnamefont{Berends}},
  \bibnamefont{and} \bibinfo{author}{\bibfnamefont{R.}~\bibnamefont{van
  Gulik}}, \bibinfo{journal}{Nucl. Phys. B} \textbf{\bibinfo{volume}{523}},
  \bibinfo{pages}{423} (\bibinfo{year}{1998}), \eprint{hep-ph/9710462}.

\bibitem[{\citenamefont{Leutgeb and Rebhan}(2020)}]{Leutgeb:2019gbz}
\bibinfo{author}{\bibfnamefont{J.}~\bibnamefont{Leutgeb}} \bibnamefont{and}
  \bibinfo{author}{\bibfnamefont{A.}~\bibnamefont{Rebhan}},
  \bibinfo{journal}{Phys. Rev. D} \textbf{\bibinfo{volume}{101}},
  \bibinfo{pages}{114015} (\bibinfo{year}{2020}), \eprint{1912.01596}.

\bibitem[{\citenamefont{Cappiello et~al.}(2020)\citenamefont{Cappiello, Cat\`a,
  D'Ambrosio, Greynat, and Iyer}}]{Cappiello:2019hwh}
\bibinfo{author}{\bibfnamefont{L.}~\bibnamefont{Cappiello}},
  \bibinfo{author}{\bibfnamefont{O.}~\bibnamefont{Cat\`a}},
  \bibinfo{author}{\bibfnamefont{G.}~\bibnamefont{D'Ambrosio}},
  \bibinfo{author}{\bibfnamefont{D.}~\bibnamefont{Greynat}}, \bibnamefont{and}
  \bibinfo{author}{\bibfnamefont{A.}~\bibnamefont{Iyer}},
  \bibinfo{journal}{Phys. Rev. D} \textbf{\bibinfo{volume}{102}},
  \bibinfo{pages}{016009} (\bibinfo{year}{2020}), \eprint{1912.02779}.

\bibitem[{\citenamefont{Knecht}(2020)}]{Knecht:2020xyr}
\bibinfo{author}{\bibfnamefont{M.}~\bibnamefont{Knecht}},
  \bibinfo{journal}{JHEP} \textbf{\bibinfo{volume}{08}}, \bibinfo{pages}{056}
  (\bibinfo{year}{2020}), \eprint{2005.09929}.

\bibitem[{\citenamefont{Masjuan et~al.}(2022)\citenamefont{Masjuan, Roig, and
  S{\'a}nchez-Puertas}}]{Masjuan:2020jsf}
\bibinfo{author}{\bibfnamefont{P.}~\bibnamefont{Masjuan}},
  \bibinfo{author}{\bibfnamefont{P.}~\bibnamefont{Roig}}, \bibnamefont{and}
  \bibinfo{author}{\bibfnamefont{P.}~\bibnamefont{S{\'a}nchez-Puertas}},
  \bibinfo{journal}{J. Phys. G} \textbf{\bibinfo{volume}{49}},
  \bibinfo{pages}{015002} (\bibinfo{year}{2022}), \eprint{2005.11761}.

\bibitem[{\citenamefont{L\"udtke and Procura}(2020)}]{Ludtke:2020moa}
\bibinfo{author}{\bibfnamefont{J.}~\bibnamefont{L\"udtke}} \bibnamefont{and}
  \bibinfo{author}{\bibfnamefont{M.}~\bibnamefont{Procura}},
  \bibinfo{journal}{Eur. Phys. J. C} \textbf{\bibinfo{volume}{80}},
  \bibinfo{pages}{1108} (\bibinfo{year}{2020}), \eprint{2006.00007}.

\bibitem[{\citenamefont{Colangelo
  et~al.}(2021{\natexlab{b}})\citenamefont{Colangelo, Hagelstein, Hoferichter,
  Laub, and Stoffer}}]{Colangelo:2021nkr}
\bibinfo{author}{\bibfnamefont{G.}~\bibnamefont{Colangelo}},
  \bibinfo{author}{\bibfnamefont{F.}~\bibnamefont{Hagelstein}},
  \bibinfo{author}{\bibfnamefont{M.}~\bibnamefont{Hoferichter}},
  \bibinfo{author}{\bibfnamefont{L.}~\bibnamefont{Laub}}, \bibnamefont{and}
  \bibinfo{author}{\bibfnamefont{P.}~\bibnamefont{Stoffer}},
  \bibinfo{journal}{Eur. Phys. J. C} \textbf{\bibinfo{volume}{81}},
  \bibinfo{pages}{702} (\bibinfo{year}{2021}{\natexlab{b}}),
  \eprint{2106.13222}.

\bibitem[{\citenamefont{Eichmann et~al.}(2024)\citenamefont{Eichmann, Fischer,
  Haeuser, and Regenfelder}}]{Eichmann:2024glq}
\bibinfo{author}{\bibfnamefont{G.}~\bibnamefont{Eichmann}},
  \bibinfo{author}{\bibfnamefont{C.~S.} \bibnamefont{Fischer}},
  \bibinfo{author}{\bibfnamefont{T.}~\bibnamefont{Haeuser}}, \bibnamefont{and}
  \bibinfo{author}{\bibfnamefont{O.}~\bibnamefont{Regenfelder}}
  (\bibinfo{year}{2024}), \eprint{2411.05652}.

\bibitem[{\citenamefont{Herren and Steinhauser}(2018)}]{Herren:2017osy}
\bibinfo{author}{\bibfnamefont{F.}~\bibnamefont{Herren}} \bibnamefont{and}
  \bibinfo{author}{\bibfnamefont{M.}~\bibnamefont{Steinhauser}},
  \bibinfo{journal}{Comput. Phys. Commun.} \textbf{\bibinfo{volume}{224}},
  \bibinfo{pages}{333} (\bibinfo{year}{2018}), \eprint{1703.03751}.

\bibitem[{\citenamefont{Chetyrkin et~al.}(2000)\citenamefont{Chetyrkin,
  K{\"u}hn, and Steinhauser}}]{Chetyrkin:2000yt}
\bibinfo{author}{\bibfnamefont{K.~G.} \bibnamefont{Chetyrkin}},
  \bibinfo{author}{\bibfnamefont{J.~H.} \bibnamefont{K{\"u}hn}},
  \bibnamefont{and}
  \bibinfo{author}{\bibfnamefont{M.}~\bibnamefont{Steinhauser}},
  \bibinfo{journal}{Comput. Phys. Commun.} \textbf{\bibinfo{volume}{133}},
  \bibinfo{pages}{43} (\bibinfo{year}{2000}), \eprint{hep-ph/0004189}.

\bibitem[{\citenamefont{Vainshtein}(2003)}]{Vainshtein:2002nv}
\bibinfo{author}{\bibfnamefont{A.}~\bibnamefont{Vainshtein}},
  \bibinfo{journal}{Phys. Lett. B} \textbf{\bibinfo{volume}{569}},
  \bibinfo{pages}{187} (\bibinfo{year}{2003}), \eprint{hep-ph/0212231}.

\bibitem[{\citenamefont{Knecht et~al.}(2004)\citenamefont{Knecht, Peris,
  Perrottet, and de~Rafael}}]{Knecht:2003xy}
\bibinfo{author}{\bibfnamefont{M.}~\bibnamefont{Knecht}},
  \bibinfo{author}{\bibfnamefont{S.}~\bibnamefont{Peris}},
  \bibinfo{author}{\bibfnamefont{M.}~\bibnamefont{Perrottet}},
  \bibnamefont{and}
  \bibinfo{author}{\bibfnamefont{E.}~\bibnamefont{de~Rafael}},
  \bibinfo{journal}{JHEP} \textbf{\bibinfo{volume}{03}}, \bibinfo{pages}{035}
  (\bibinfo{year}{2004}), \eprint{hep-ph/0311100}.

\bibitem[{\citenamefont{Blum et~al.}(2020)\citenamefont{Blum, Christ, Hayakawa,
  Izubuchi, Jin, Jung, and Lehner}}]{Blum:2019ugy}
\bibinfo{author}{\bibfnamefont{T.}~\bibnamefont{Blum}},
  \bibinfo{author}{\bibfnamefont{N.}~\bibnamefont{Christ}},
  \bibinfo{author}{\bibfnamefont{M.}~\bibnamefont{Hayakawa}},
  \bibinfo{author}{\bibfnamefont{T.}~\bibnamefont{Izubuchi}},
  \bibinfo{author}{\bibfnamefont{L.}~\bibnamefont{Jin}},
  \bibinfo{author}{\bibfnamefont{C.}~\bibnamefont{Jung}}, \bibnamefont{and}
  \bibinfo{author}{\bibfnamefont{C.}~\bibnamefont{Lehner}}
  (\bibinfo{collaboration}{RBC, UKQCD}), \bibinfo{journal}{Phys. Rev. Lett.}
  \textbf{\bibinfo{volume}{124}}, \bibinfo{pages}{132002}
  (\bibinfo{year}{2020}), \eprint{1911.08123}.

\bibitem[{\citenamefont{Chao et~al.}(2021)\citenamefont{Chao, Hudspith,
  G\'erardin, Green, Meyer, and Ottnad}}]{Chao:2021tvp}
\bibinfo{author}{\bibfnamefont{E.-H.} \bibnamefont{Chao}},
  \bibinfo{author}{\bibfnamefont{R.~J.} \bibnamefont{Hudspith}},
  \bibinfo{author}{\bibfnamefont{A.}~\bibnamefont{G\'erardin}},
  \bibinfo{author}{\bibfnamefont{J.~R.} \bibnamefont{Green}},
  \bibinfo{author}{\bibfnamefont{H.~B.} \bibnamefont{Meyer}}, \bibnamefont{and}
  \bibinfo{author}{\bibfnamefont{K.}~\bibnamefont{Ottnad}},
  \bibinfo{journal}{Eur. Phys. J. C} \textbf{\bibinfo{volume}{81}},
  \bibinfo{pages}{651} (\bibinfo{year}{2021}), \eprint{2104.02632}.

\bibitem[{\citenamefont{Chao et~al.}(2022)\citenamefont{Chao, Hudspith,
  G\'erardin, Green, and Meyer}}]{Chao:2022xzg}
\bibinfo{author}{\bibfnamefont{E.-H.} \bibnamefont{Chao}},
  \bibinfo{author}{\bibfnamefont{R.~J.} \bibnamefont{Hudspith}},
  \bibinfo{author}{\bibfnamefont{A.}~\bibnamefont{G\'erardin}},
  \bibinfo{author}{\bibfnamefont{J.~R.} \bibnamefont{Green}}, \bibnamefont{and}
  \bibinfo{author}{\bibfnamefont{H.~B.} \bibnamefont{Meyer}},
  \bibinfo{journal}{Eur. Phys. J. C} \textbf{\bibinfo{volume}{82}},
  \bibinfo{pages}{664} (\bibinfo{year}{2022}), \eprint{2204.08844}.

\bibitem[{\citenamefont{Blum et~al.}(2025)\citenamefont{Blum, Christ, Hayakawa,
  Izubuchi, Jin, Jung, Lehner, and Tu}}]{Blum:2023vlm}
\bibinfo{author}{\bibfnamefont{T.}~\bibnamefont{Blum}},
  \bibinfo{author}{\bibfnamefont{N.}~\bibnamefont{Christ}},
  \bibinfo{author}{\bibfnamefont{M.}~\bibnamefont{Hayakawa}},
  \bibinfo{author}{\bibfnamefont{T.}~\bibnamefont{Izubuchi}},
  \bibinfo{author}{\bibfnamefont{L.}~\bibnamefont{Jin}},
  \bibinfo{author}{\bibfnamefont{C.}~\bibnamefont{Jung}},
  \bibinfo{author}{\bibfnamefont{C.}~\bibnamefont{Lehner}}, \bibnamefont{and}
  \bibinfo{author}{\bibfnamefont{C.}~\bibnamefont{Tu}}
  (\bibinfo{collaboration}{RBC, UKQCD}), \bibinfo{journal}{Phys. Rev. D}
  \textbf{\bibinfo{volume}{111}}, \bibinfo{pages}{014501}
  (\bibinfo{year}{2025}), \eprint{2304.04423}.

\bibitem[{\citenamefont{Fodor et~al.}(2024)\citenamefont{Fodor, G{\'e}rardin,
  Lellouch, Szab\'o, Toth, and Zimmermann}}]{Fodor:2024jyn}
\bibinfo{author}{\bibfnamefont{Z.}~\bibnamefont{Fodor}},
  \bibinfo{author}{\bibfnamefont{A.}~\bibnamefont{G{\'e}rardin}},
  \bibinfo{author}{\bibfnamefont{L.}~\bibnamefont{Lellouch}},
  \bibinfo{author}{\bibfnamefont{K.~K.} \bibnamefont{Szab\'o}},
  \bibinfo{author}{\bibfnamefont{B.~C.} \bibnamefont{Toth}}, \bibnamefont{and}
  \bibinfo{author}{\bibfnamefont{C.}~\bibnamefont{Zimmermann}}
  (\bibinfo{collaboration}{BMWc}) (\bibinfo{year}{2024}), \eprint{2411.11719}.

\bibitem[{\citenamefont{Prades et~al.}(2009)\citenamefont{Prades, de~Rafael,
  and Vainshtein}}]{Prades:2009tw}
\bibinfo{author}{\bibfnamefont{J.}~\bibnamefont{Prades}},
  \bibinfo{author}{\bibfnamefont{E.}~\bibnamefont{de~Rafael}},
  \bibnamefont{and}
  \bibinfo{author}{\bibfnamefont{A.}~\bibnamefont{Vainshtein}},
  \bibinfo{journal}{Adv. Ser. Direct. High Energy Phys.}
  \textbf{\bibinfo{volume}{20}}, \bibinfo{pages}{303} (\bibinfo{year}{2009}),
  \eprint{0901.0306}.

\bibitem[{\citenamefont{Redmer}(2024)}]{Redmer:2024bva}
\bibinfo{author}{\bibfnamefont{C.~F.} \bibnamefont{Redmer}},
  \bibinfo{journal}{Nuovo Cim. C} \textbf{\bibinfo{volume}{47}},
  \bibinfo{pages}{247} (\bibinfo{year}{2024}).

\bibitem[{\citenamefont{Ablikim et~al.}(2020)}]{BESIII:2020nme}
\bibinfo{author}{\bibfnamefont{M.}~\bibnamefont{Ablikim}} \bibnamefont{et~al.}
  (\bibinfo{collaboration}{BESIII}), \bibinfo{journal}{Chin. Phys. C}
  \textbf{\bibinfo{volume}{44}}, \bibinfo{pages}{040001}
  (\bibinfo{year}{2020}), \eprint{1912.05983}.

\bibitem[{\citenamefont{Altmannshofer et~al.}(2019)}]{Belle-II:2018jsg}
\bibinfo{author}{\bibfnamefont{W.}~\bibnamefont{Altmannshofer}}
  \bibnamefont{et~al.} (\bibinfo{collaboration}{Belle-II}),
  \bibinfo{journal}{PTEP} \textbf{\bibinfo{volume}{2019}},
  \bibinfo{pages}{123C01} (\bibinfo{year}{2019}), \bibinfo{note}{[Erratum: PTEP
  {\bf 2020}, 029201 (2020)]}, \eprint{1808.10567}.

\end{thebibliography}

\end{document}